\documentclass[12pt]{article}
\usepackage[utf8]{inputenc}
\usepackage{graphicx}
\graphicspath{Figures/}
\usepackage{amsmath}
\usepackage{amssymb}
\usepackage{listings}
\usepackage{hyperref}
\usepackage{cuted}
\usepackage{authblk}
\usepackage{threeparttable}
\usepackage{tikz}
\usepackage{setspace}
\usepackage{float}
\usepackage{lineno}
\usepackage{xspace}
\usepackage{colortbl}
\usepackage{pdflscape}
\usepackage{longtable}
\usepackage[authoryear]{natbib}

\usepackage{bbm}
\newcommand{\bbeta}{ \mbox{\boldmath $\beta$}}

\newcommand{\bX}{ \mbox{\bf X}}

\newcommand{\bY}{ \mbox{\bf Y}}

\newcommand{\bM}{ \mbox{\bf M}}
\newcommand{\bW}{ \mbox{\bf W}}

\newcommand{\iid}{\stackrel{iid}{\sim}}

\newcommand{\beq}{ \begin{equation}}
\newcommand{\eeq}{ \end{equation}}
\newcommand{\beqn}{ \begin{eqnarray}}
\newcommand{\eeqn}{ \end{eqnarray}}

\usepackage{caption}
\title{\vspace{-4em} Mediation analysis of community context effects on heart failure using the survival R2D2 prior}
\author[1,3]{Brandon R. Feng}
\author[2]{Eric Yanchenko}
\author[4]{K. Lloyd Hill}
\author[5]{Lindsey A. Rosman}
\author[1]{Brian J. Reich}
\author[3]{Ana G. Rappold}
\affil[1]{Department of Statistics, North Carolina State University}
\affil[2]{Global Connectivity Program, Akita International University}
\affil[3]{United States Environmental Protection Agency, Center for Public Health and Environmental Assessment}
\affil[4]{Oak Ridge Associated Universities at United States Environmental Protection Agency, Center for Public Health and Environmental Assessment}
\affil[5]{Department of Medicine, Division of Cardiology, University of North Carolina at Chapel Hill}

\begin{document}

\maketitle
\vspace{-2em}
\begin{abstract}
\singlespacing
        Congestive heart failure (CHF) is a leading cause of morbidity, mortality and healthcare costs, impacting $>$23 million individuals worldwide. Large electronic health records data provide an opportunity to improve clinical management of diseases, but statistical inference on large amounts of relevant personal data is still challenging. Thus, accurately identifying influential risk factors is pivotal to reducing information dimensionality. Bayesian variable selection in survival regression is a common approach towards solving this problem. Here, we propose placing a beta prior directly on the model coefficient of determination (Bayesian $R^2$), which induces a prior on the global variance of the predictors and provides shrinkage. Through reparameterization using an auxiliary variable, we are able to update a majority of the parameters with Gibbs sampling, simplifying computation and quickening convergence. Performance gains over competing variable selection methods are showcased through an extensive simulation study. Finally, the method is applied in a mediation analysis to identify community context attributes impacting time to first congestive heart failure diagnosis of patients enrolled in University of North Carolina Cardiovascular Device Surveillance Registry. The model has high predictive performance and we find that factors associated with higher socioeconomic inequality increase risk of heart failure.
        
    % With the increasing amount of available covariates in medical data, identification of the most influential factors is pivotal in survival regression modeling. Bayesian analyses focusing on variable selection are a common approach towards solving this problem. However, most use MCMC methods that require many tuning parameters which may slow convergence. In this paper, we propose placing a beta prior directly on the model coefficient of determination (Bayesian $R^2$), which induces a prior on the global variance of the predictors and provides shrinkage. Through reparameterization using an auxiliary variable, we are able to update a majority of the parameters with sequential Gibbs sampling, simplifying computation and quickening convergence. Performance gains over competing variable selection methods are then showcased through an extensive simulation study. Finally, the method is applied to identify attributes in the community context impacting time to congestive heart failure of patients enrolled in University of North Carolina Cardiovascular Device Surveillance Registry. The model has high predictive performance with a C-index of over 0.7 and we find factors associated with higher socioeconomic inequality increase risk of heart failure.

\end{abstract}


\newpage

\section{Introduction}

Congestive heart failure (CHF) is a progressive condition, impacting $>$23 million individuals worldwide \citep{roger2013epidemiology} and is associated with a gradual worsening in quality of life, decreased physical function, frequent hospitalizations, increased healthcare costs and high rates of premature death. The majority of causes that lead to CHF are attributed to non-modifiable factors such as genetics, race and gender \citep{bozkurt2024hf}. However, a consequential portion of burden of disease is influenced by potentially intervenable factors such as occupation, lifestyle and environmental factors \citep{james2015review, dadvand2016green, bhatnagar2017environmental, chen2024deep}. Understanding the role of community context on disease burden provides an opportunity to both quantify the impact of these potentially intervenable set of risk factors and understand their effect on total health burden and creation of disparities between patients. 

 ``Community context" encompasses social and environmental factors such as neighborhood walkability, pollution levels and socioeconomic indicators which are common risk exposures that can similarly affect large portions of the world population. The number of studies related to the effect of community context on CHF has grown in recent years with various areas of interest. \cite{malambo2016built} focused on the role of physical activity and transportation on CHF. The links between socioeconomic status and demographic information as risk factors for heart failure were analyzed by \cite{lawson2020risk}. Finally, \cite{shah2013global} and \cite{jia2023effect} analyzed the impact of different sources of air pollution on risk of heart failure. Since these risks in community context are common to many, once correctly identified, they can be used to better model individual time-to-event across large populations.

Growth of electronic health records and large environmental databases provide an unprecedented opportunity to quantify the impacts of the environment on incidence and progression of CHF. The dimension of the data necessitates development of new statistical techniques for identifying risk measures in both parametric and non-parametric inference. Recently, machine learning methods have been leveraged to predict CHF events and identify the most predictive community context risk factors \citep{chen2024deep}. For example, \cite{guo2021comprehensive} and \cite{ wang2021machine} compared multiple predictive models on CHF risk based on personal characteristics such as age, sex, comorbidities and medications. \cite{guo2021comprehensive} focused on feature importance from the XGBoost model while \cite{wang2021machine} reported selected features from both tree and regression models. In addition, \cite{miao2018predictive} proposed a novel split rule for accurate feature selection in regression tree methods to predict CHF with high performance metrics. Most of the statistical models are focused on improving predictive performance, however, rather than inference for risk factors. 

High-dimensional variable selection has been widely studied ranging from frequentist penalized regression methods \citep{tibshirani1996regression, hoerl1970ridge, zou2005regularization} to Bayesian methods including spike and slab \citep{rovckova2018spike, nie2023bayesian}, empirical Bayes \citep{bar2020scalable, scott2010bayes} and Global-Local shrinkage priors \citep{carvalho2009handling, bhattacharya2015dirichlet, zhang2022bayesian}. Global-Local methods shrink individual coefficients to drastically reduce coefficient estimation error in comparison to penalties that enforce a global shrinkage across all terms. Recently, the R2D2 prior has shown promise as a global-local shrinkage prior. First introduced by \cite{zhang2022bayesian} for the normal linear model and extended to generalized linear models by \cite{yanchenko2023spatial, yanchenko2021r2d2}, the R2D2 prior places a prior directly on the coefficient of determination, $R^2$. It has been proven to have higher prior mass near 0 with heavier tails than competing Global-Local methodology to allow for increased shrinkage and less estimation bias than its competitors such as LASSO, Horseshoe and Dirichlet-Laplace \citep{zhang2022bayesian}. In survival analysis and right-censored regression modeling contexts, Cox proportional hazards models with variable selection \citep{tibshirani1997lasso, goeman2010l1, gui2005penalized, lee2011bayesian, mu2021bayesian} in both frequentist and Bayesian settings are widely utilized. Another popular framework is the accelerated failure time model where many penalized regression techniques have been developed for both Weibull regression and more generalized variants \citep{ahmed2012lasso, lee2017variable, rockova2012hierarchical, newcombe2017weibull, liang2023adaptive}. However, most proposed samplers in Bayesian variable selection for Weibull regression do not utilize Gibbs sampling and are sensitive to tuning parameter selection for candidate distributions and acceptance probabilities.

%Survival time-to-event applications range from various health event outcomes \citep{liu2023using, swindell2009accelerated, zhang2016parametric, cavalcante2023weibull} to different areas of system reliability analysis \citep{smith1991weibull, compare2020industrial}. 
%In this work, we propose  

%ocus on health outcomes with application to heart failure similar to \citep{ahmad2017survival, panahiazar2015using}. %
%With growing amounts of medical data both publicly and privately available, there is increased importance placed in the ability to select the most important covariates for regression analysis.

In this paper, we propose an R2D2 prior for right-censored Weibull models. The proposed model builds upon the Weibull model, first introduced in \cite{yanchenko2021r2d2}, to include the impact of censoring and provides a novel nearly fully Gibbs MCMC sampling scheme for faster convergence and reduction of tuning parameters. We demonstrate the utility of the proposed R2D2 variable selection method for survival models in identifying mediators to the total effect of community context on the development of CHF. Mediator variables were selected from electronic health records and daily pacemaker reports in adult patients with implanted pacemaker device enrolled in the University of North Carolina Cardiovascular Device Surveillance Registry (UNC CDSR) \citep{serang2017exploratory}.

The paper is structured as follows. Section 2 introduces both the Weibull regression and R2D2 frameworks. Section 3 outlines the MCMC sampler. Section 4 tests R2D2 against competing variable selection techniques in a comprehensive simulation study. Finally, in Section 5, the method is applied to find important mediators to analyze the impact of community context on time to CHF through a mediation analysis. Derivations are provided in the Appendix.

\section{Methodology}

 For observation $i \in \{1,...,n\}$, define $Y_i$ as the survival time and $\bX_i = (X_{i1},...,X_{ip})^T$ as the corresponding covariates. Observation $i$ is censored if $Y_i$ exceeds censoring time $T_i$, with $\delta_i = I(Y_i>T_i)$ serving as the binary censoring indicator. 
We assume the survival times follow a Weibull distribution where $Y_i|\theta,\beta_0,\bbeta \sim \mbox{Weibull}\{\theta, \exp(\eta_i)\}$, for shape parameter $\theta$ and log-scale parameter
$$\eta_i = \beta_0 + \bX_i^T\bbeta,$$
so that
 $\beta_0$ is the intercept, $\bbeta = (\beta_1,...,\beta_p)^T$ and $\beta_j$ is the coefficient of the $j^{th}$ covariate. We note that the Weibull model applies to both the accelerated-failure-time and proportional-hazards setting and the method described below can be extended to a flexible scale mixture of Weibull distributions that spans a wide class of survival distributions (Supplementary Section 1). The Weibull distribution is parameterized to have density function 

 \begin{equation} \label{eq: StandardWeibull}
        f_Y(y_i|\theta, \beta_0, \boldsymbol{\beta}) = \theta y_i^{\theta-1} e^{-\theta \eta_i} e^{-y_i^\theta e^{-\theta \eta_i}}.
\end{equation}
 
  The variable selection prior for the regression coefficients is $\beta_j|\phi_j, W \iid \mbox{Normal}(0, \phi_j W)$, where $W>0$ represents the global variance and $\phi_j>0$ represents local shrinkage for covariate $j$.  The local shrinkage factors are given constraint $\sum_{j=1}^p \phi_j = 1$ so that $\phi_j$ represents the proportion of variance allocated to covariate $j$. During computation, $\mathbf{X}$ is scaled with respect to the mean and standard deviation of the uncensored observations.

\subsection{Dirichlet Decomposition}

    The variance parameters $\phi_j$ reflect the relative amount of variance appropriated to each model component. In global penalties such as Ridge and LASSO \citep{van2019shrinkage}, these parameters are fixed and the same penalty is placed across all covariates. For global-local priors such as the Horseshoe \citep{carvalho2009handling} and Dirichlet-Laplace \citep{bhattacharya2015dirichlet}, these $\phi_j$'s will have a prior distribution placed on them to adaptively shirnk different covariate effects. A common prior \citep{bhattacharya2015dirichlet, zhang2022bayesian} is $\boldsymbol{\phi} \sim \text{Dirichlet}(\xi_1,...,\xi_p)$ which satisfies $\sum_{j=1}^p \phi_j = 1$ and allows a straightforward interpretation of each covariate's variance as a percentage of the total variance. The parameters $\xi_1,...,\xi_p$ are often set to be equal to some $\xi_0$, i.e., $\xi_1 =...=\xi_p = \xi_0$. Large values of $\xi_0$ sets $\phi_j$ to be all approximately equal to $\frac{1}{p}$ while small values of $\xi_0$ encourage sparsity with some $\phi_j$, and thus $\beta_j$, near zero, and some large $\phi_j$ for important covariates \citep{zhang2022bayesian, yanchenko2023spatial, yanchenko2021r2d2}.

\subsection{Survival R2D2 Prior}

Here, we introduce a prior distribution based on the Bayesian coefficient of determination ($R^2$) defined by \cite{gelman2019r}. We follow a similar framework as \cite{zhang2022bayesian} and \cite{yanchenko2023spatial, yanchenko2021r2d2}. Define the mean and variance functions $\mu(\eta)$ and $\sigma^2(\eta)$ where $E(Y|\eta) = \mu(\eta)$ and $V(Y|\eta) = \sigma^2(\eta)$ and recall $\eta = \bX \boldsymbol{\beta}$. The general Bayesian $R^2$ is defined as 
\begin{equation}\label{eq:R2_def}
    R^2(W) = \frac{V\{\mu(\eta)\}}{V\{\mu(\eta)\} + E\{\sigma^2(\eta)\}},
\end{equation} 
where the expectations in \eqref{eq:R2_def} are with respect to $\boldsymbol{\beta}$ given $\beta_0$ and $W$.
 We will denote this as $R^2$ for the rest of the paper and suppress its dependencies. As a more complex model can explain more variation in the data, this $R^2$ definition holds as a measure of model fit.

If $Y|\eta, \theta \sim Weibull(\theta, e^{\eta})$, then $\mu(\eta) = e^{\eta}\Gamma(1+\frac{1}{\theta})$ and $\sigma^2(\eta) = e^{2\eta}[\Gamma(1+\frac{2}{\theta}) - \{\Gamma(1+\frac{1}{\theta})\}^2]$. The $R^2$ is derived as (see Appendix A.1): 

\begin{equation} \label{eq:prior_R2}
    R^2 = \frac{(e^{W} - 1)}{\frac{\Gamma(1+\frac{2}{\theta})}{\{\Gamma(1+\frac{1}{\theta})\}^2}e^W -1}.
\end{equation}

\noindent $R^2$ is bounded below by $R^2_{min} = 0$ and above by $R^2_{max} = \frac{\{\Gamma(1+\frac{1}{\theta})\}^2}{\Gamma(1+\frac{2}{\theta})}$. Censoring is ignored as an artifact of data collection, which should not affect the choice of prior. Even though the observed data may be censored, the underlying process is still a Weibull distribution. 

A Beta prior \citep{zhang2022bayesian, yanchenko2021r2d2} is typically placed on $R^2$ such that $R^2 \sim \text{Beta}(a,b)$ as it is highly flexible with the desired support. However, the support of $R^2$, [$R^2_{min}, R^2_{max}$], is not [0,1] and we instead induce a $\text{Beta}(a,b)$ prior on the shifted and scaled $\Tilde{R}^2 = \frac{R^2 - R^2_{min}}{R^2_{max} - R^2_{min}}$ \citep{yanchenko2021r2d2}. Inducing a $\text{Beta}(a,b)$ on $\Tilde{R}^2$ is equivalent to inducing a four-parameter Beta distribution on the original $R^2$ such that $R^2 \sim \text{Beta}(a,b,R^2_{min},R^2_{max})$. From \eqref{eq:R2_def}, there exists a one-to-one transform of the global variance term, $W$, to $R^2$. Therefore, placing a prior on $W$ induces a beta prior on $R^2$.

\begin{figure}[ht!]
    \centering
    \includegraphics[width=1\linewidth, scale = 1.5]{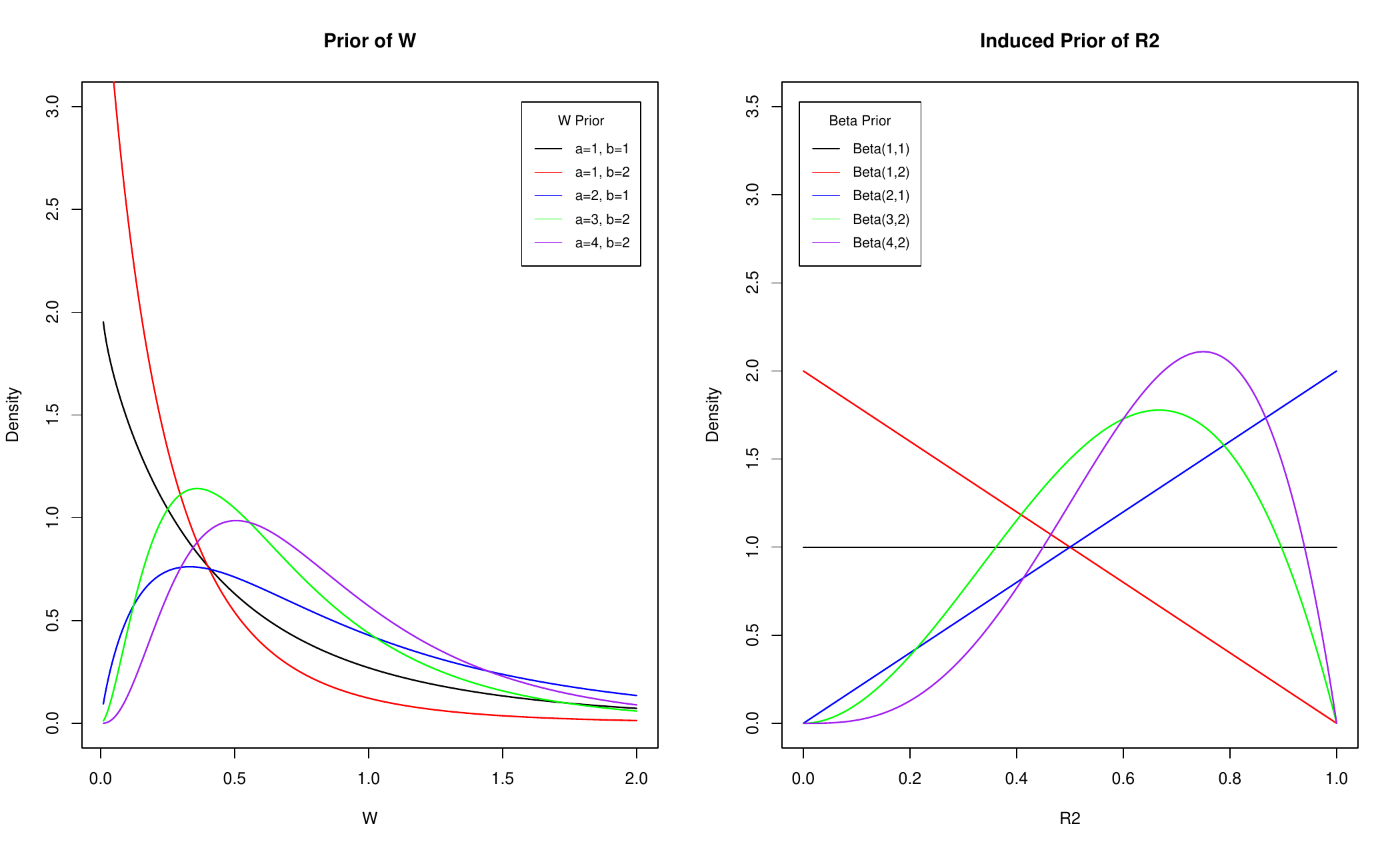}
    \caption{Various priors on $W$ and the subsequent induced $\text{Beta}(a,b)$ prior.}
    \label{fig:WtoR2}
 \end{figure}

 $\Tilde{R}^2 \sim \text{Beta}(a,b,0,R^2_{max})$ induces the prior for $W$ defined as 

\begin{equation} \label{eq:prior_w}
    \pi(w|\theta) = \frac{1}{B(a,b)}\frac{e^W c^a |d||d-c|(e^W-1)^{a-1}(-d^2 + cd)^{b-1}}{d^b(e^Wc-d)^{a+b}}
\end{equation}

\noindent where $c = \Gamma(1+\frac{2}{\theta})$ and $ d = \{\Gamma(1+\frac{1}{\theta})\}^2$. We select values of $a$ and $b$ for the prior of $W$ based on our assumption of model fit. For $a \ll b$, we assume a poor model fit with the prior mean $R^2$ and for $a \gg b$, the converse occurs where we assume a good model fit. Poor model fit will correspond to smaller values of $W$ and induce higher sparsity while good model fit will have larger values of $W$ and assume more covariates as significant. This is illustrated in Figure \ref{fig:WtoR2}. 

\subsubsection{Generalized Beta Prime Approximation}

\begin{figure}[ht!]
    \centering
    \includegraphics[width=1\linewidth, scale = 1.5]{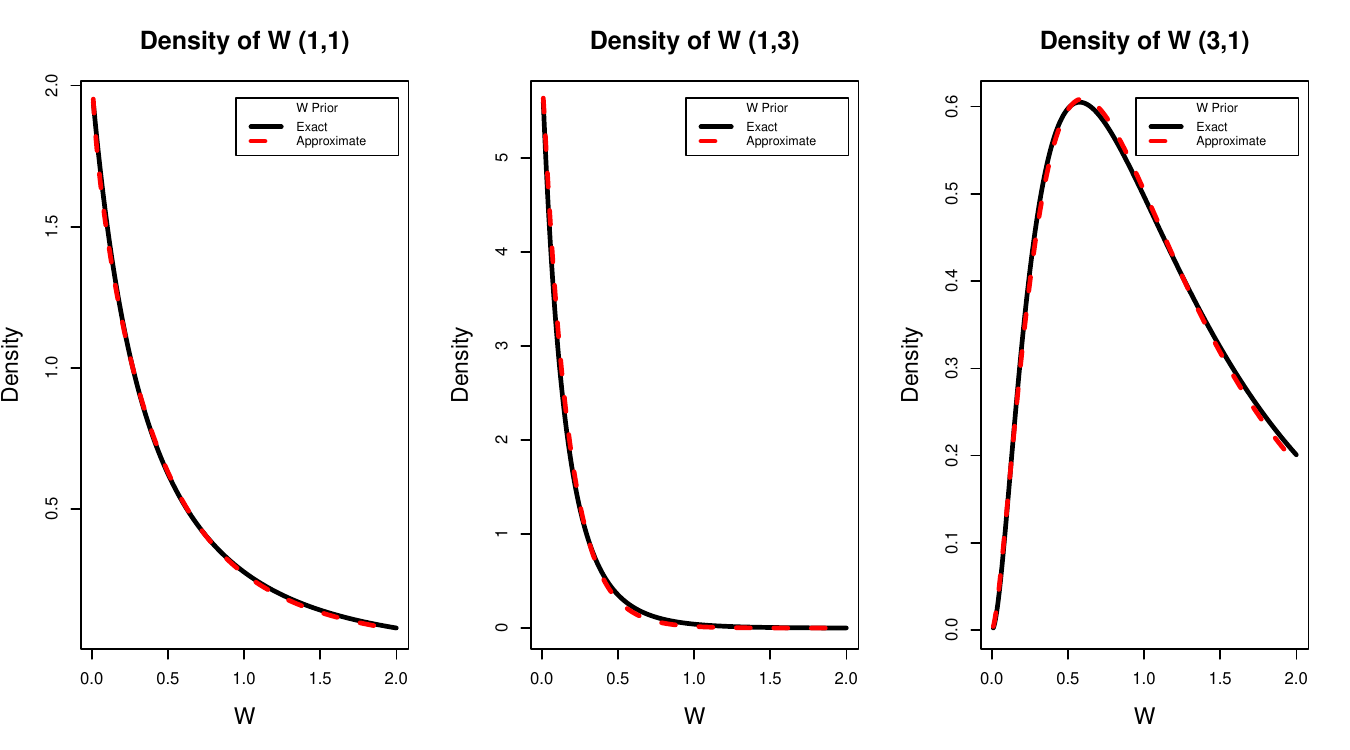}
    \caption{The exact density of various priors on $W$ with the $GBP$ approximation}
    \label{fig:W_Approx}
 \end{figure}

    The prior on W is neither a standard distribution nor conjugate. The Generalized Beta Prime (GBP) distribution is a flexible four parameter distribution that allows for a close approximation of many more complex distributions. We follow the method raised by \cite{yanchenko2021r2d2} to approximate the prior of $W$ with this GBP distribution so $\pi(w) \sim GBP(a^*, b^*, c^*, d^*)$. The Pearson $\chi^2$-divergence is minimized between the true and approximate PDF of $W$ through an optimization procedure on the parameters of the $GBP$. We find that the approximation is robust for various choices of $a \text{ and } b$ on the original prior of $W$ in Figure \ref{fig:W_Approx}. Holding the value of $c^* = 1$ allows for the approximation to be sampled through a fully Gibbs approach, making convergence faster and our methodology more computationally efficient. The prior on $W$ is thus $W\sim GBP(a^*,b^*,1,d^*)$. It is not computationally feasible to re-estimate these parameters at each iteration, thus we find $a^*,b^*,d^*$ from the MLE estimate of $\theta$. 
    
    We take advantage of the fact that $W \sim GBP(a^*, b^*, 1, d^*)$ can be reparameterized further as $W|\xi \sim Gamma(a^*, \gamma)$, $\gamma \sim Gamma(b^*, d^*)$. The full model specification with priors is as follows:
    
     \begin{align*} 
        Y_i|\eta_i, \theta &\sim Weibull(\theta, e^{\beta_0 + X_i\boldsymbol{\beta}}) \\ 
        \beta_j|\phi_j, W &\sim N(0, \phi_j W)\text{, } j \in \{1,...,p\} \\
        \phi_1,...,\phi_p &\sim Dir(\xi_1,...,\xi_p) \tag{\theequation}\label{eq: Model_prior}\\ 
        W|\gamma &\sim Gamma(a^*, \gamma) \\
        \gamma & \sim Gamma(b^*, d^*) \\
        \log(\theta) \sim &N(t_1, t_2)\text{, } \beta_0 \sim N(\mu_{\beta_0}, \sigma^2_{\beta_0}) 
    \end{align*}

    \noindent with hyperparameters $t_1, t_2, \mu_{\beta_0}, \sigma^2_{\beta_0}$ selected to give uninformative priors.

\section{Posterior Computation}

 With a large number of parameters, an efficient sampler with fast convergence is needed for feasible computation on larger datasets. The posterior for the model specified in (\ref{eq: Model_prior}) is approximated with a combination of Gibbs and Metropolis-Hastings sampling where the intercept and shape parameters are drawn through Metropolis-Hastings sampling while all others are drawn through Gibbs sampling. The posterior of the $\beta$'s are calculated closely following \cite{damlen1999gibbs}. 

\subsection{Weibull Reparameterization}

    The addition of a auxiliary variable into the Weibull distribution by \cite{damlen1999gibbs} is essential to allowing for Gibbs sampling on the $\beta$ coefficients. We show this derivation for our context here. The censored likelihood function is
    
    \begin{equation} \label{eq:WeibLike}
        f(\bold{Y}|\theta, \boldsymbol{\beta}) = \prod_{i=1}^n \theta^{\delta_i} y_i^{\delta_i(\theta-1)} e^{-\theta \delta_i X_i\boldsymbol{\beta}} e^{-y_i^\theta e^{-\theta X_i\boldsymbol{\beta}}}.
    \end{equation}

    \noindent Assuming a prior $\pi(\boldsymbol{\beta}) \sim N(0, \Sigma)$ where $\Sigma = \text{Diag}(\boldsymbol{\phi})W$,

    \begin{equation} \label{eq:BetaLike}
        f(\boldsymbol{\beta}|\bold{Y}, \theta,  \boldsymbol{\phi}, W) \propto \prod_{i=1}^ n \{e^{-y_i^\theta e^{-\theta X_i\beta}}\} \cdot e^{-\frac{1}{2}\beta^T\Sigma^{-1}\beta -\theta \sum_{i=1}^n \delta_i x_i\beta}.
    \end{equation}

    \noindent Introducing a latent truncated exponential random variable $\bold{U} = (u_1,...,u_n)$ allows for the conditional distribution of each $u_i$ to be independent truncated exponential \citep{damlen1999gibbs}. This likelihood then becomes

    \begin{equation} \label{eq:BetaULike}
        f(\boldsymbol{\beta}, \bold{U}|\theta, \bold{Y}, \boldsymbol{\phi}, W) \propto \prod_{i=1}^n \{e^{-u_i} I(u_i > y_i^\theta e^{-\theta x_i\boldsymbol{\beta}})\}\cdot e^{-\frac{1}{2}\boldsymbol{\beta}^T\Sigma^{-1}\boldsymbol{\beta} -\theta \sum_{i=1}^n \delta_i x_i\boldsymbol{\beta}}.
    \end{equation}

    \noindent For the $i^{th}$ observation, marginalizing $u_i$ returns:
    \begin{equation} \label{eq:UMargin}
    \begin{split}
    \int f(\boldsymbol{\beta},u_i|\theta, y_i, x_i, \boldsymbol{\phi}, W) \,du_i &\propto e^{-\frac{1}{2}\boldsymbol{\beta}^T\Sigma^{-1}\boldsymbol{\beta}-\theta x_i\boldsymbol{\beta}} \int \{e^{-u_i} I(u_i > y_i^\theta e^{-\theta x_i\boldsymbol{\beta}})\}\, du_i \\ &= e^{-y_i^\theta e^{-\theta x_i\boldsymbol{\beta}}} \cdot e^{-\frac{1}{2}\boldsymbol{\beta}^T\Sigma^{-1}\boldsymbol{\beta} -\theta x_i\boldsymbol{\beta}},
    \end{split}
    \end{equation}
    \noindent which equals \eqref{eq:BetaLike} as desired. The likelihood transformation allows for conjugate updates for $\boldsymbol{U}$ and $\boldsymbol{\beta}$ and becomes the basis of our following MCMC posterior derivations on $\boldsymbol{\beta}$.

\subsection{MCMC Steps}

The MCMC algorithm is as follows:

\begin{enumerate}
    \item $\beta_0|\bold{Y}, \bold{X}, \boldsymbol{\beta}, \theta, \boldsymbol{\phi}, W \sim $ Metropolis-Hastings step

    \item For $j \in \{1,...,p\}$:
    
    $\beta_j|\bold{Y}, \bold{U}, \beta_0, \boldsymbol{\beta}_{(j)}, \theta, \boldsymbol{\phi}, W \sim N(-\theta \phi_jW \Tilde{n} \bar{X}^*_j, \phi_jW)$ 
    
    truncated to 
    $
    \begin{cases}
    \beta_j > -\frac{\log(u_i) - \theta \log(y_i) + \theta \sum_{l\neq j}X_{i,l}\beta_l}{ \theta X_{i,j}} & X_{i,j} > 0\\
    \beta_j < -\frac{\log(u_i) - \theta \log(y_i) + \theta \sum_{l\neq j}X_{i,l}\beta_l}{\theta X_{i,j}} & X_{i,j} < 0\\
    \text{Undefined} &  X_{i,j} = 0
    \end{cases}
    $

    where $\Tilde{n}$ is the number of uncensored observations, $\bar{X}^*_j$ is the mean of uncensored observations which is equal to 0 after scaling.

    \item $\theta|\bold{Y}, \bold{X}, \boldsymbol{\beta} \sim$ Metropolis-Hastings step

    \item $W|\boldsymbol{\beta}, \boldsymbol{\phi}, \gamma \sim GIG(\sum_{j=1}^p \frac{\beta_j^2}{\phi_j}, 2\gamma, a^* - \frac{p}{2})$

    \item $\gamma|W \sim Gamma(a^*+b^*, d^* + W)$
     
    \item For $j \in \{1,...,p\}:$

    $T_j|\beta_j, \gamma \sim GIG(\beta_j^2, 2\gamma, \frac{a^*}{p} - \frac{1}{2})$

    $\phi_j = \frac{T_j}{\sum_{j=1}^p T_j}$

    \item For $i \in \{1,...,n\}:$

    $u_i|y_i, x_i, \beta_0, \boldsymbol{\beta}, \theta \sim Exp(1)I\{u_i > y_i^\theta e^{-\theta(\beta_0 + X_i\boldsymbol{\beta})}\}$
     
\end{enumerate}

Although $\theta$ and $\beta_0$ have conjugate priors, they are difficult to implement and computationally unstable. Thus, we use Metropolis-Hastings to update both. A log reparameterization is applied to make the candidate distribution of $\theta$ Normal. The acceptance ratio is tuned during the burn-in stage to be between 30\% and 50\%. The truncated normal aspect of this algorithm can be computationally unstable for cases where the mean is not centered at 0. Thus, we used the specific scaling scheme and are unable to have the intercept be a Gibbs update. For full posterior derivations, please see the Appendix.

\section{Simulation Study}

In this section, we compare R2D2 with competing Bayesian variable selection priors in the Weibull regression setting to show performance increases in key metrics for estimation error and selection accuracy. Let $p \in \{ 100, 500 \}$ with the number of uncensored observations set to $n=60$ for $p=100$ and $n = 100$ for $p = 500$ to reflect a high-dimensional setting. The covariates $X_{i1},...,X_{ip}$ are generated from an $AR(1)$ structure with correlation $\rho = 0.5$ or $\rho = 0.9$ and then standardized as described in Section 2. $\beta_0$ is randomly generated from a Normal(0, 1) distribution. Let $\boldsymbol{\beta}_1 = (2.5, -2, 0.5, -1, 1.5)$ and $\boldsymbol{\beta}_2 = (-1.5, -0.5, 2, -2.5, 1)$. For $p = 100$, the coefficients are $\boldsymbol{\beta} = \{\boldsymbol{0}_5, \boldsymbol{\beta}_1, \boldsymbol{0}_5, \boldsymbol{\beta}_2, \boldsymbol{0}_{80} \}$. For $p = 500$, the coefficients are $\boldsymbol{\beta} = \{ \boldsymbol{0}_5, \boldsymbol{\beta}_1, \boldsymbol{0}_5, \boldsymbol{\beta}_2, \boldsymbol{0}_{5}, -\boldsymbol{\beta}_1, \boldsymbol{0}_5, - \boldsymbol{\beta}_2, \boldsymbol{0}_{460}\}$. 35\% of the generated $Y_i$'s are censored through fixed right censoring all observations above the $65^{th}$ percentile of the $Y_i$'s. For each setting combination, 100 datasets are simulated from the Weibull distribution outlined in (\ref{eq: StandardWeibull}) with a true $\log(\theta) = 0.5$. The true Bayesian $R^2$ is approximately 0.7.

Competing methods are the Horseshoe prior \citep{carvalho2009handling}, Ridge penalized Cox regression \citep{hoerl1970ridge, verweij1994penalized} and LASSO penalized Cox regression \citep{tibshirani1996regression, tibshirani1997lasso}. The Horseshoe prior is implemented using our Gibbs sampler and the Cox regressions are implemented from the glmnet package with the penalty $\lambda$ selected using the one standard error method \citep{hastie2009elements}. We showcase 5 different prior settings for R2D2 with (a,b) representing the parameters of a Beta(a,b) prior. The first is a (0.5,0.5) prior that is uninformative on model fit. The second is a (1,5) prior that assumes the covariates do not explain much variation and penalize towards a simpler model. The third is a (5,1) prior that assumes there are many significant covariates for a more complex model. The fourth and fifth priors are more extreme variants of the previous two with (0.5, 4) and (4, 0.5) priors. For the Bayesian estimates, MCMC is run for 100,000 samples with the first 30,000 treated as burn-in and thinning every 3 samples.

Tables \ref{tab:10cov0.5MSE} and \ref{tab:10cov0.9MSE} report the overall average sum of squared error (SSE) of $\boldsymbol{\beta}$, the overall average SSE of non-zero $\boldsymbol{\beta}$'s and the overall average SSE of $\boldsymbol{\beta}$'s with true value zero. For the Bayesian methods, the point estimate is the posterior median. For the Cox regressions, the Cox $\boldsymbol{\beta}$'s are transformed by negating them and multiplying them by the true $\theta$ to be analogous to the Weibull $\boldsymbol{\beta}$'s.

\begin{table*}[htbp]
    \centering
    \caption{ $\boldsymbol{\rho}$ \textbf{= 0.5}. Average sum of squared error (SSE) for all $\boldsymbol{\beta}$'s, non-zero $\boldsymbol{\beta}$'s and zero $\boldsymbol{\beta}$'s from 100 simulated datasets. Standard errors are noted in parenthesis. The highest performing value in each category is highlighted.}
    \label{tab:10cov0.5MSE}%
    \resizebox{\textwidth}{!}{\begin{tabular}{ |c||c|c|c||c|c|c| }
        \hline
        & & p = 100 & & & p = 500 &\\ [2pt]
        \hline
        \textbf{Prior}& Overall & Non-zero & Zero & Overall & Non-zero & Zero\\ [2pt]
       \hline
          R2D2 (0.5,0.5) & \textbf{0.347 (0.025)} & \textbf{0.317 (0.024)} & 0.030 (0.004) & \textbf{1.148 (0.071)} & 0.850 (0.055) & 0.299 (0.029) \\  
          R2D2 (1,5) & 2.062 (0.445) & 2.044 (0.445) & 0.018 (0.005) & 28.325 (2.650) & 28.185 (2.664) & 0.140 (0.029) \\  
          R2D2 (0.5,4) & 0.986 (0.138) & 0.964 (0.137) & 0.022 (0.005) & 25.729 (2.628) & 25.573 (2.642) & 0.157 (0.026) \\  
          R2D2 (5,1) & 0.401 (0.029) & 0.356 (0.027) & 0.045 (0.006) & 1.460 (0.084) & 0.911 (0.054) & 0.549 (0.043) \\ 
          R2D2 (4,0.5) & 0.410 (0.043) & 0.358 (0.033) & 0.052 (0.014) & 1.294 (0.069) & \textbf{0.836 (0.046)} & 0.458 (0.039) \\ 
          HS & 0.517 (0.042) & 0.382 (0.026) & 0.136 (0.028) & 1.547 (0.130) & 1.125 (0.103) & 0.423 (0.037) \\ 
          Ridge & 27.383 (0.025) & 27.382 (0.025) & \textbf{0.002 (0.001)} & 54.785 (0.040) & 54.778 (0.042) & \textbf{0.007 (0.002)} \\ 
          LASSO & 5.604 (0.703) & 5.194 (0.625) & 0.410 (0.098) & 24.455 (0.951) & 23.657 (0.985) & 0.798 (0.063) \\ 
       \hline
    \end{tabular}}
\end{table*}

\begin{table*}[htbp]
    \centering
    \caption{ $\boldsymbol{\rho}$ \textbf{= 0.9}. Average sum of squared error (SSE) for all $\boldsymbol{\beta}$'s, significant $\boldsymbol{\beta}$'s and non-significant $\boldsymbol{\beta}$'s from 100 simulated datasets. Standard errors are noted in parenthesis. The highest performing value in each category is highlighted.}
    \label{tab:10cov0.9MSE}%
    \resizebox{\textwidth}{!}{\begin{tabular}{ |c||c|c|c||c|c|c| }
        \hline
        & & p = 100 & & & p = 500 &\\ [2pt]
        \hline
        \textbf{Prior}& Overall & Non-zero & Zero & Overall & Non-zero & Zero \\ [2pt]
        \hline
          R2D2 (0.5,0.5) & \textbf{3.840 (0.357)} & \textbf{3.807 (0.359)} & 0.032 (0.015) & \textbf{9.894 (0.966)} & \textbf{9.793 (0.961)} & 0.100 (0.022) \\ 
          R2D2 (1,5) & 22.918 (0.254) & 22.876 (0.249) & 0.041 (0.015) & 45.971 (0.335) & 45.918 (0.332) & 0.053 (0.018) \\ 
          R2D2 (0.5,4) & 22.479 (0.039) & 22.434 (0.034) & 0.045 (0.015) & 45.442 (0.320) & 45.379 (0.317) & 0.062 (0.019) \\  
          R2D2 (5,1) & 5.893 (0.604) & 5.849 (0.604) & 0.044 (0.015) & 21.736 (1.587) & 21.652 (1.585) & 0.085 (0.021) \\ 
          R2D2 (4,0.5) & 4.789 (0.470) & 4.740 (0.470) & 0.049 (0.017) & 16.389 (1.352) & 16.299 (1.350) & 0.090 (0.020) \\ 
          HS & 4.306 (0.387) & 4.253 (0.388) & 0.053 (0.014) & 21.659 (1.488) & 21.581 (1.483) & 0.078 (0.021) \\ 
          Ridge & 27.359 (0.017) & 27.355 (0.018) & \textbf{0.004 (0.001)} & 54.729 (0.025) & 54.714 (0.027) & \textbf{0.015 (0.002)} \\ 
          LASSO & 17.361 (2.330) & 14.062 (1.393) & 3.299 (1.079) & 48.636 (0.264) & 48.536 (0.272) & 0.100 (0.019) \\ 
       \hline
    \end{tabular}}
\end{table*}

\newpage 

From Tables \ref{tab:10cov0.5MSE} and \ref{tab:10cov0.9MSE}, it is clear that R2D2 outperforms the competing methods in all three cases. For the $p = 100$ case, the lowest error is always from one of the R2D2 priors and the (0.5,0.5) prior almost always performs the best across both autocorrelation settings. This difference in error is more apparent in the more challenging $p = 500$ case. For $\rho = 0.9$, the SSE of R2D2 is around half of Horseshoe's and under a fifth of Ridge and LASSO.

Tables \ref{tab:10cov0.5AUC} and \ref{tab:10cov0.9AUC} compare the methods in terms of selecting the correct subset of variables. For the Bayesian methods, covariate $j$ is deemed significant if the 95\% credible interval of $\beta_j$ excludes zero. For LASSO regression, it is deemed significant if the estimated coefficient is nonzero. Accuracy of variable selection is compared using the area under the Receiver-Operating Characteristic curve (AUC). Posterior coverage is the Bayesian analogy to 95\% confidence intervals where we record the proportion of the 100 simulated 95\% credible intervals that include the true $\beta$ value. 

Tables \ref{tab:10cov0.5AUC} and \ref{tab:10cov0.9AUC} also evaluate prediction. The average in-sample concordance index (C-index) to show predictive power is also included. The C-index is analogous to AUC for survival outcome prediction \citep{pencina2004overall}. Higher C-index values indicate more accurate survival prediction with a value of 0.5 equivalent to random guessing of the survival outcome. We calculate the C-index based off of the posterior medians of the $\beta$'s and $\theta$ for each Bayesian model. For the Cox model, we take the C-index corresponding to the selected $\lambda_{1se}$ for the ridge and LASSO penalties by the cv.glmnet function.

\begin{table*}[htbp]
    \centering
    \caption{ $\boldsymbol{\rho}$ \textbf{= 0.5}. Average area under the Receiver-Operating Characteristic curve (AUC), posterior coverage and C-index from 100 simulated datasets. Standard errors are noted in parenthesis. The highest performing value in each category is highlighted.}
    \label{tab:10cov0.5AUC}%
    \resizebox{\textwidth}{!}{\begin{tabular}{ |c||c|c|c||c|c|c| }
        \hline
        & & p = 100 & & & p = 500 &\\ [2pt]
        \hline
        \textbf{Prior}& AUC & Coverage & C-index & AUC & Coverage & C-index \\ [2pt]
        \hline
          R2D2 (0.5,0.5) & \textbf{0.993 (0.001)} & \textbf{0.993 (0.001)} & 0.950 (0.001) & \textbf{0.970 (0.006)} & \textbf{0.988 (0.001)} & 0.972 (0.001) \\ 
          R2D2 (1,5) & 0.889 (0.030) & 0.963 (0.003) & 0.915 (0.014) & 0.525 (0.049) & 0.973 (0.002) & 0.854 (0.014) \\ 
          R2D2 (0.5,4) & 0.989 (0.001) & 0.971 (0.002) & 0.945 (0.002) & 0.522 (0.049) & 0.975 (0.001) & 0.856 (0.015) \\ 
          R2D2 (5,1) & 0.992 (0.001) & 0.992 (0.001) & 0.951 (0.001) & 0.942 (0.007) & 0.981 (0.002) & 0.977 (0.001) \\ 
          R2D2 (4,0.5) & 0.992 (0.001) & 0.992 (0.001) & 0.951 (0.001) & 0.945 (0.007) & 0.983 (0.002) & 0.975 (0.001) \\  
          HS & 0.980 (0.005) & 0.985 (0.003) & \textbf{0.961 (0.001)} & 0.952 (0.006) & 0.982 (0.001) & \textbf{0.978 (0.001)} \\ 
          Ridge & NA & NA & 0.665 (0.005) & NA & NA & 0.605 (0.004) \\ 
          LASSO & 0.693 (0.006) & NA & 0.909 (0.002) & 0.623 (0.005) & NA & 0.831 (0.004) \\  
        \hline
    \end{tabular}}
\end{table*}

\begin{table*}[htbp]
    \centering
    \caption{ $\boldsymbol{\rho}$ \textbf{= 0.9}. Average area under the Receiver-Operating Characteristic curve (AUC), posterior coverage and C-index from 100 simulated datasets. Standard errors are noted in parenthesis. The highest performing value in each category is highlighted.}
    \label{tab:10cov0.9AUC}%
    \resizebox{\textwidth}{!}{\begin{tabular}{ |c||c|c|c||c|c|c| }
        \hline
        & & p = 100 & & & p = 500 &\\ [2pt]
        \hline
        \textbf{Prior}& AUC & Coverage & C-index & AUC & Coverage & C-index \\ [2pt]
        \hline
          R2D2 (0.5,0.5) & \textbf{0.967 (0.010)} & \textbf{0.970 (0.002)} & 0.880 (0.003) & \textbf{0.968 (0.014)} & \textbf{0.984 (0.001)} & \textbf{0.895 (0.008)} \\ 
          R2D2 (1,5) & 0.473 (0.048) & 0.908 (0.001) & 0.581 (0.024) & 0.687 (0.045) & 0.962 (0.000) & 0.532 (0.028) \\ 
          R2D2 (0.5,4) & 0.531 (0.048) & 0.909 (0.001) & 0.558 (0.025) & 0.736 (0.043) & 0.963 (0.000) & 0.529 (0.028) \\
          R2D2 (5,1) & 0.925 (0.021) & 0.961 (0.002) & 0.860 (0.010) & 0.938 (0.022) & 0.977 (0.001) & 0.759 (0.026) \\ 
          R2D2 (4,0.5) & 0.955 (0.014) & 0.966 (0.002) & 0.877 (0.003) & 0.946 (0.020) & 0.981 (0.001) & 0.813 (0.022) \\ 
          HS & 0.965 (0.098) & 0.969 (0.019) & \textbf{0.888 (0.027)} & 0.967 (0.139) & 0.977 (0.008) & 0.748 (0.278) \\ 
          Ridge & NA & NA & 0.682 (0.005) & NA & NA & 0.663 (0.004) \\  
          LASSO & 0.663 (0.011) & NA & 0.825 (0.004) & 0.789 (0.012) & NA & 0.777 (0.003) \\
       \hline
    \end{tabular}}
\end{table*}

We can see from Tables \ref{tab:10cov0.5AUC} and \ref{tab:10cov0.9AUC} that R2D2 outperforms the other methods in the categories related to variable selection in both settings of $p$. In AUC and coverage, R2D2 (0.5,0.5) is the top performer with small standard errors, indicating stability of results. In the $p = 100$ setting, the Bayesian methods achieve over 95\% coverage from the 95\% credible intervals calculated in the low autocorrelation case and the higher performing methods also surpass 95\% coverage in the high case. For the C-index, although R2D2 does not have the highest value we see high predictive power in the (0.5,0.5), (5,1) and (4,0.5) cases that are comparable to Horseshoe and higher than Ridge and LASSO. In the $p = 500$ setting, we see R2D2 attain the highest AUC and Coverage across both values of $\rho$ with the C-index being the highest as well in the $\rho = 0.9$ case. 

We believe that R2D2 (0.5,0.5) performs the best from the flexibility this prior definition brings. The (0.5,0.5) prior has a horseshoe-like shape allows for high mass at both very simple and complex models. This allows for the distribution of $W$ to allow for both levels of sparsity. The more sparse (1,5) and (0.5,4) priors penalize too much towards a simple model, leading to lower coefficient error on the zero coefficients, but also over-penalizing the non-zero. The (5,1) and (4,0.5) priors allow for more non-zero coefficients, however, they under-select with non-significant coefficients being left in the model.

\section{Application: Effects of Community Context on Incident Heart Failure }

%Congestive heart failure (CHF) is a prevalent CVD with over 5.8 million cases in the United States and 23 million worldwide \citep{roger2013epidemiology}. The estimated annual cost of CHF in the US is \$30.7 billion with a global cost of \$108.1 billion \citep{cook2014annual}. Thus, it is of interest to identify and characterize impact of modifiable risk factors such as built and social environment. In this section, we demonstrate a utility of the proposed variable selection method to identify direct and indirect effects of community context on development of CHF. More specifically, we use R2D2 model to identify a set of risk factors that mediate total effect of community context on progression of CVD and diagnosis of CHF. 

In this section, we demonstrate the utility of the proposed variable selection method to identify the effect of community context variables on development of "incident CHF". We use data from a cohort of patients $>$18 years of age ($n=2,577$) enrolled in the University of North Carolina (UNC) Cardiovascular Device Surveillance Registry (UNC CDSR).  The UNC CDSR \citep{rosman2021arrhythmia, rosman2022immediate} is an ongoing prospective, clinical research registry of patients who have received implanted cardiac devices at the UNC Medical Center and 10 affiliated hospitals located throughout central North Carolina. The registry collects daily device data from all remote monitoring transmissions and routine follow‐up clinic visits for all cardiac device implants, upgrades, and replacements.Selected patients were restricted to those who had received a pacemaker between 2010 and 2021 and had continuous device data throughout the study period.Device data were deterministically linked to patient-level sociodemographics, clinical and medication history which were routinely extracted from electronic health records (EHR) using standard procedures and validated International Classification of Diseases (ICD)-9 and ICD-10 codes. Data from the UNC CDSR have been used extensively in research and clinical care \citep{rosman2024wearable, rosman2024patient, rosman2022immediate, rosman2021arrhythmia, cleland2023impact, mazzella2022effects}.

The primary outcome of interest $(\bY)$ was defined as time (years) from device implant to the first incident of CHF diagnosis. Within the cohort, 1,212 (47\%) experienced CHF during the follow up period. Clinical diagnoses were considered present if the ICD-9 or ICD-10 code for that specific condition were recorded during hospitalization or in at least two outpatient encounters. Because the patients have a device and followed over time, time to first diagnosis is unlikely to be confounded with access to care. This labeling is used extensively in research conducted with Medicare and other claims data, and has been shown to enhance diagnostic accuracy in these data sources \citep{tirschwell2002validating}. 

Community context variables $(\bX)$ were contextualized based on $p = 72$ factors which included socioeconomic indicators, pollution levels, area level demographics, and vegetation type. Community context data were extracted from several different sources and linked to patients based on their geocoded residential address. For example, land cover metrics (e.g., tree canopy cover, urban impervious surface cover) were averaged within 1-km circular buffers drawn around a patient's geocoded address. Census-based demographic data were matched to patients based on their census tract of residence. To reduce the impact of multicollinearity among community context variables, we used principal component decomposition to identify orthogonal components among community context factors. The standardized projection onto the first $p_x = 14$ eigenvectors, $\bX^* = \bX \bW$, where $\bW$ is the $p \times p_x$ matrix of eigenvectors explaining 70\% of total variability, was used as the multivariate treatment.

There are multiple mechanisms by which an individual develops CHF \citep{bozkurt2024hf}. Patients with a pacemaker have an underlying cardiovascular condition which led them to receive their device and often have one or more risk factors for developing CHF. Many of the risk factors can be considered as potential mediators of the total effect of community context on the progression of disease and first diagnosis of CHF. Mediators to CHF are important in this context because they provide insight in how the total effect of community characteristics is linked with respect to existing risk factors. It is of interest to know whether community context is directly linked to the incidence of CHF, or if other risk factors play a mediating role. 

In total, 49 clinically relevant individual-level covariates $(\bM)$ including race, age, lifestyle factors, medical comorbidities, and medications were selected as potential mediators of the total effect of community context on time to first CHF diagnosis. The statistical challenge was to identify a subset of $\bM$ that may mediate the effect of community context. We fit $\bM$ onto $\bY$ and apply R2D2 to select a subset of mediators with a significant effect on the outcome of interest similar to \cite{luo2020high}. We also fit a LASSO penalized Cox model with the $\lambda$ penalty selected through the one standard error method. The C-index is used as the performance metric for comparison. A summary of the dataset can be found in Section 3 of the Supplementary.

%For the individual-level covariates, comorbidities and medications are binary indicator variables with 1 representing condition existence prior to device implantation and 0 otherwise.  

% To show the Weibull model is appropriate model for this dataset, a Weibull regression is fit where the response is days to CHF and the predictors are the 49 individual level covariates and the first 30 community context principal components. The survival function of the fitted model residuals is plotted against a Kaplan-Meier estimate of the ordered residuals. As shown in \ref{fig:Resid_KM}, the Weibull model is a close approximation of the Kaplan-Meier curve and the study proceeds assuming this is an appropriate distributional assumption.

% \begin{figure}[ht!]
%     \centering
%     \includegraphics[width=1\linewidth, scale = 1.5]{PaperFigures/AFT_KM.pdf}
%     \caption{The survival function of Weibull model residuals plotted against the Kaplan-Meier estimate of the residuals.}
%     \label{fig:Resid_KM}
%  \end{figure}
 
\subsection{R2D2 for Mediator Selection}

%More specifically,  hypertension, atrial fibrillation and dyslipidemia were selected as significant comorbidity; diuretics, anticoagulant and calcium channel blockers were selected as significant medications; while mean hours of daily physical activity was selected as a significant lifestyle related mediator. 

%$$ log(Y_i) = \beta_0 + \sum_{i}$$
The first step taken to estimate direct and indirect effects of community context variables was to identify which risk factors may mediate the effect. From the initial 49 individual-level potential mediator factors, R2D2 selected $p_M = 7$ variables as significant, marking them as potential mediators of the total effect: prescribed diuretics, anticoagulants and calcium channel blockers, have atrial fibrillation, hypertension, dyslipidemia and average time spent in physical activity (measured in minutes per day by the device) after cohort admission (see Table \ref{tab: SigCov} in Appendix \ref{Appendix: MedEffect}). Coefficient interpretation follows that of the accelerated failure time model. For example, time to incident CHF diagnosis was 23\% shorter for patients with dyslipidemia than those without ($100\%*(1-e^{-0.26}) \approx 23\%$). This means that having been diagnosed with dyslipidemia decreases the time to first CHF diagnosis, hence increases the risk. The in-sample C-index was found to be 0.720 and out-of-sample is 0.706, indicating high predictive performance.

In comparison, LASSO found 19 covariates as significant with an in-sample C-index of 0.716 and out-of-sample 0.706. The covariates selected by R2D2 were a subset of those selected by LASSO. R2D2 selected approximately three times fewer covariates, allowing for better interpretability with similar predictive performance.

\subsection{Mediated Effect of Community Context}

 In the second step,  we used R2D2 selected mediator variables to quantify direct and indirect effect of community context on development of CHF.   
 %in a multiple mediaton analysis \citep{vanderweele2014mediation}. 
 More specifically, we 
 fit non-penalized Bayesian regression outcome and mediator models to obtain the posterior distribution of the indirect effect \citep{vanderweele2014mediation}\citep{yuan2009bayesian}. For comparison, the results are compared to the assessment based on the LASSO variable selection.  %The multiple mediation methodology applied is detailed as follows. 
 
 For patient $i \in \{1,...,n\}$, given time to CHF $Y_i$, indicator $M_{j,i}$ for $j \in \{1,...,p_M\}$ where $p_M$ is the number of individual-level mediators previously selected, and community context projection $X^*_{k,i}$ for $k \in \{1,...,p_X\}$, to estimate direct effects, we fit overall Weibull outcome model
$$ Y_i|\theta_0, \xi_0, \boldsymbol{\beta}, \boldsymbol{\tau}^D \sim \text{Weibull}\{\theta_0, \exp(\xi_0 + \sum_{j=1}^{p_M} \beta_j M_{j,i} + \sum_{k=1}^{p_X} \tau^D_k X^*_{k,i})\}$$

%$$Y_i|\theta_0, \xi,  \tau^* \sim \text{Weibull}\{\theta, \exp(\xi + \sum_{k=1}^{14} \tau^*_k X^*_{k,i})\}, $$
\noindent where $\tau^D_k$ represents the direct effect of the $k^{th}$ projection in $\bX_i^*$. 

To capture indirect effects, for the $j^{th}$ mediator, we model
%while $\tau^*_k$ represents the total effect of the $k^{th}$ projection in $\bX_i^*$. 
$$
\begin{cases} 
    M_{j,i} = \xi_j + \sum_{k=1}^{14} \alpha_{k,j} X^*_{k,i} + \epsilon_j & \text{if $M_j$ is continuous} \\
    \log \,\{Pr(M_{j,i} = 1)\} = \xi_j + \sum_{k=1}^{14} \alpha_{k,j} X^*_{k,i}  & \text{if $M_j$ is binary}
\end{cases}
$$

\noindent where $\epsilon_j \sim N(0,1)$ \citep{vanderweele2011causal,vanderweele2016mediation}. Let $c_j = I(M_j \text{ is continuous})$ serve as the indicator if the $j^{th}$ mediator is continuous. The indirect effect of the $k^{th}$ projection on outcome $Y_i$ is measured as:
\begin{equation} \label{eq: IE_k}
    \tau^I_k = \sum_{j=1}^{p_M} \alpha_{k,j} \beta_j \left[ c_j + \biggl\{\frac{1}{n} \sum_{i=1}^n \frac{\exp(\xi_j +  \alpha_{k,j} X^*_{k,i})}{(1 + \exp(\xi_j + \alpha_{k,j} X^*_{k,i}))^2}\biggl\}(1-c_j) \right].
\end{equation}

%$X_i^*$ is also directly regressed onto $Y_i$ to model the total effect of community context on time to CHF. The Weibull model is 

 \noindent The estimate of the binary mediator indirect effect was derived by \cite{li2007estimation}. Let $\boldsymbol{\tau}^I$ represent the vector of $\tau^I_k \text{ for }k \in \{1,...,p_X\}$. The indirect effects of the original community context variables, $\boldsymbol{\omega}$, are calculated through multiplying $\boldsymbol{\tau}^I$  by the rotation matrix $\bW$ where $\boldsymbol{\omega} = \bW\boldsymbol{\tau}^I$. The full derivation of this can be found in Appendix \ref{Appendix: IndEffect}. Total effects were calculated by adding the indirect effect and direct effect of each community context covariate at each iteration.

 %Indirect effects from the LASSO mediators generally had bounds closer to 0, indicating the additional LASSO mediators may be inflating the precision and selecting too many indirect effects as significant. This strengthens the need for accurate mediator selection.

  Tables \ref{tab: MultMed_R2D2} and \ref{tab: MultMed_LASSO} summarize the association between community context variables and the time to first incidence of CHF for significant total effects calculated using R2D2 and LASSO selected mediators, respectively. The tables provide means and 95\% posterior credible intervals for total, direct and indirect effects expressed as the expected difference (added or reduced) in number of days to first incidence associated with one standard deviation increase in each community context variable. Given community context coefficients $\omega_i, \text{ } i \in \{1,...,72\}$ and average cohort age, difference in days to CHF caused by the $i^{th}$ community context factor, $\Delta_{chf}$, is calculated as 
 $$\Delta_{chf} = (e^{\omega_i} - 1)\cdot \bar{\text{Age}} \cdot 365.$$ 
 
\begin{figure}[ht!]
    \centering
    \includegraphics[width=1\linewidth, scale = 1.0]{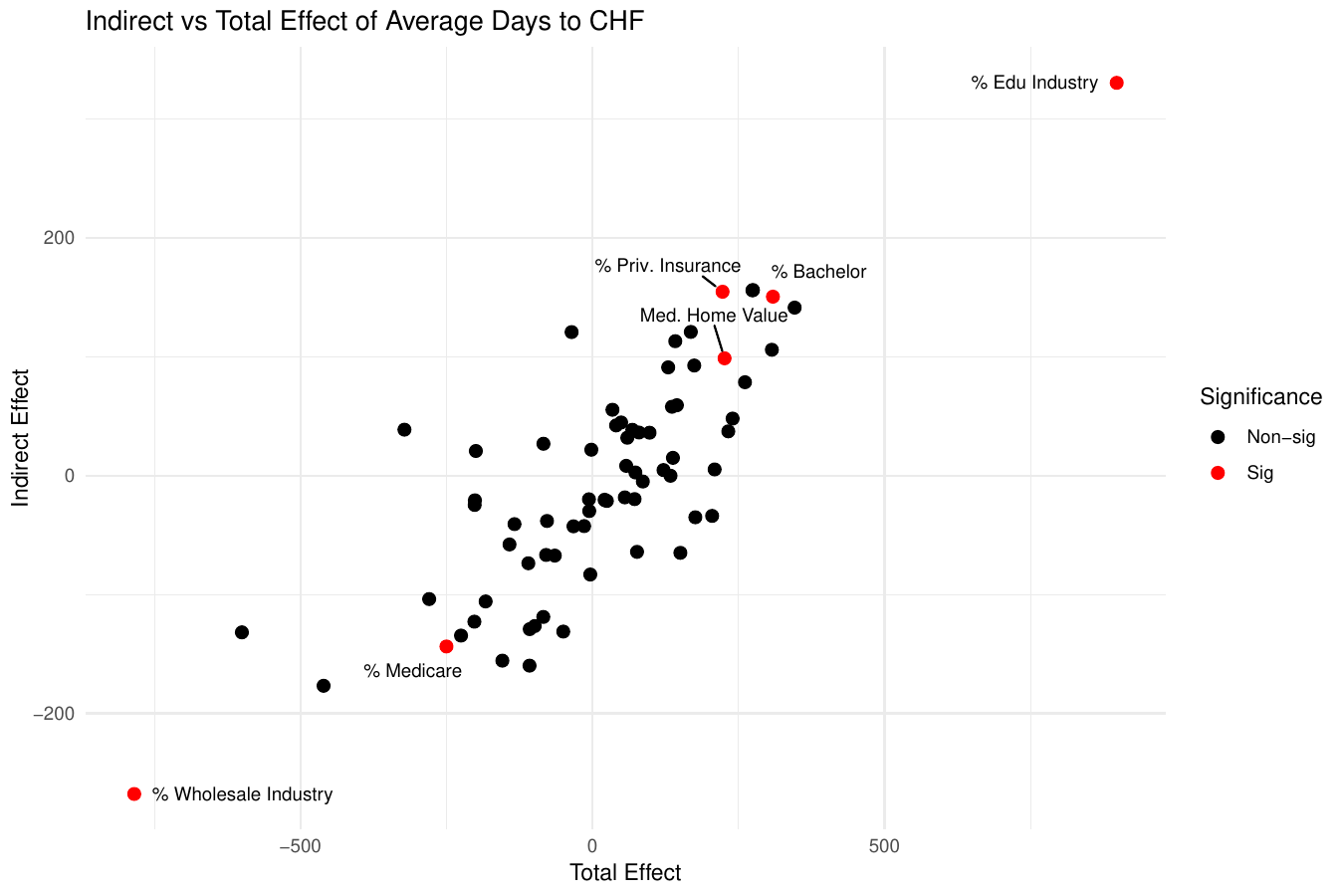}
    \caption{Posterior means of the difference in days to CHF resulting from total and indirect effect sizes of community context factors mediated by R2D2 selected mediators. Red points represent significant total effects.}
    \label{fig:R2D2_Tot_IEffect}
 \end{figure}

  For indirect effects, 11 mediated through R2D2 mediators were significant as opposed to 14 for LASSO mediators. Similarly, for direct effects, R2D2 mediators found 1 as significant while LASSO mediators found 5. However, both analysis from R2D2 and LASSO mediators found the same 6 total effects as significant, making the results from R2D2 more parsimonious. For total effects, statistically significant increase in time to CHF was observed for \% employed in higher education, \% of people with bachelors degree or higher, median home value and \% of people with private insurance. Meanwhile, statistically significant association and decreased time to CHF was observed for \% of people with medicare insurance and \% employed in the wholesale trade industry. The average proportion of indirect effects to these total effects, or proportion mediated \cite{vanderweele2013policy}, was around 48\% for R2D2 and 45\% for LASSO selected mediators. Figure \ref{fig:R2D2_Tot_IEffect} shows the posterior mean effect size of community context total effects against that of indirect effects mediated through R2D2 mediators with significant total effects highlighted.
 
 The significant factors that extended time to development of CHF were associated with more affluent neighborhoods. For example, from the R2D2 mediated values, each increase in standard deviation of median home value adds around 226 days to time to CHF. Meanwhile, factors that decreased time to CHF are associated with higher poverty areas. Again from the R2D2 mediated values, each increase in standard deviation in percentage with medical insurance decreases days to CHF by around 250 days. These values represent around 20\% and 22\%, respectively, of the average days to CHF after pacemaker implantation. Occupation distribution also highly impacts time to CHF. Areas with higher percentages working in higher education have longer time to CHF while those with high percentages working in the wholesale trade industry have decreased time to CHF. These estimates indicate policies towards decreasing CHF prevalence should be targeted towards areas of the community causing socioeconomic inequality.

\begin{table*}[h!]
\small
    \centering
    \caption{Posterior medians and 95\% credible intervals of the difference in average days to CHF with a 1 standard deviation change from significant total effects of community context factors for total, direct and indirect effects mediated by R2D2 selected mediators. Column 2 represents the original mean and standard deviation of community context covariates before scaling.}
    \label{tab: MultMed_R2D2}%
    \resizebox{16cm}{!}{\begin{tabular}{ |l|c|c|c|c|}
        \hline
        Covariate & Mean (SD) & Indirect Effect & Direct Effect & Total Effect\\ 
        \hline
        Wholesale Trade Industry \% & 3.1 (1.9)  & -267.8 (-480.8, -54.1) & -521.6 (-1067.8, -13.3) & -784.5 (-1348.7, -242.4) \\ 
        Medicare Insurance \% & 17.6 (8.7)& -143.7 (-225.0, -63.1) & -106.8 (-353.5, 120.9) & -250.0 (-509.1, -22.9) \\ 
        Private Insurance \% & 66.4 (15.4) & 154.8 (82.6, 228.4) & 67.6 (-120.8, 248.4) & 222.7 (22.7, 420.5) \\  
        Median Home Value & $1.8 \cdot 10^5$ ($9.9 \cdot 10^4$) & 98.8 (27.8, 172.3) & 127.0 (-37.6, 289.4) & 226.2 (48.6, 401.4) \\ 
        Bachelors \% & 31.9 (21.4) & 150.7 (83.1, 221.4) & 157.5 (-10.0, 290.3) & 309.0 (124.8, 457.9) \\  
        Higher Education Industry \% & 24.5 (7.3) & 330.6 (56.8, 613.4) & 560.6 (-79.3, 1237.8) & 897.7 (180.4, 1660.3) \\ 
       \hline
    \end{tabular}}
\end{table*}

\begin{table*}[h!]
\small
    \centering
    \caption{Posterior medians and 95\% credible intervals of the difference in average days to CHF with a 1 standard deviation change from significant total effects of community context factors for total, direct and indirect effects mediated by LASSO selected mediators. Column 2 represents the original mean and standard deviation of community context covariates before scaling.}
    \label{tab: MultMed_LASSO}%
    \resizebox{16cm}{!}{\begin{tabular}{ |l|c|c|c|c|}
        \hline
        Covariate & Mean (SD) & Indirect Effect & Direct Effect & Total Effect\\ 
        \hline
        Wholesale Trade Industry \% & 3.1 (1.9) & -236.6 (-441.9, -33.9) & -555.4 (-1012.9, -106.7) & -787.4 (-1272.2, -295.5) \\ 
        Medicare Insurance \% & 17.6 (8.7) & -136.2 (-212.7, -61.1) & -100.6 (-273.1, 81.5) & -236.3 (-428.2, -40.3) \\ 
        Private Insurance \% & 66.4 (15.4) & 154.7 (86.0, 224.8) & 55.1 (-96.0, 220.2) & 210.0 (40.7, 392.4) \\ 
        Median Home Value & $1.8 \cdot 10^5$ ($9.9 \cdot 10^4$) & 84.9 (17.2, 154.9) & 140.7 (13.9, 289.1) & 226.0 (78.5, 383.0) \\ 
        Bachelors \% & 31.9 (21.4)  & 146.0 (81.8, 213.8) & 178.0 (37.3, 337.5) & 325.0 (170.8, 495.6) \\ 
        Higher Education Industry \% & 24.5 (7.3) & 324.8 (74.3, 583.2) & 793.6 (51.4, 1581.6) & 1127.4 (352.0, 1988.2) \\ 
       \hline
    \end{tabular}}
\end{table*}

% \begin{figure}[ht!]
%     \centering
%     \includegraphics[width=1\linewidth, scale = 1.0]{PaperFigures/Mult_MedCI.pdf}
%     \caption{95\% Posterior Credible Intervals of the 10 most negative and 10 most positive multiple mediated effect sizes of community context.}
%     \label{fig:Mult_BE_IEffect}
%  \end{figure}

\clearpage
\section{Discussion}

 Accurately identifying the most influential mediators between community-level risk factors and development of CHF provides insight into the pathways by which CHF occurs. Novel and rich data sets provide an unprecedented opportunity to discover and analyze the effect of multiple pathways, but present an inferential challenge for dealing with large numbers of variables. In this paper, we introduce the global-local R2D2 prior for variable selection in survival analysis. We extend the work of \cite{yanchenko2021r2d2} to accommodate censoring in the Weibull model, as well as deriving a flexible mixture distribution for cases of model misspecification. The prior on $R^2$ allows us to incorporate a wide range of beliefs of model fit directly into variable selection. We also propose a novel MCMC sampling approach utilizing mostly Gibbs updates for faster convergence and less tuning than alternative sampling schemes. 

The method is shown to have less estimation error and higher selection accuracy than popular competing variable selection techniques through a robust simulation study showcasing various levels of sparsity, covariate correlation and dimension size. Mediation analysis is then used to analyze the effect of community context on development of chronic heart failure in a cohort of patients with pacemakers from the University of North Carolina CDSR database to identify key factors in community context impacting CHF risk. From this, we show that multiple factors that define socioeconomic and environmental community context variables associated with high socioeconomic inequality and high pollution increase the risk of CHF.

Although we find no significant spatial correlation in this dataset, we believe that the proposed method can be extended to spatial survival analysis. There also exists computational instability for extreme priors that causes convergence issues in the Beta Prime approximation of the distribution of the global W variance term. Future work can also focus on a different measure of model fit for survival models other than Bayesian $R^2$. Overall, the applications and future directions of R2D2 are numerous and have the potential to improve the analysis of data in medicine, environmental science, and beyond.

\section{Acknowledgements}

 Thanks to North Carolina State University, Oak Ridge Institute for Science and Education and the Environmental Protection Agency for supporting this work. We also thank the device clinic staff and the participants of the UNC Cardiovascular Device Surveillance Registry (UNC CDSR) for their valuable contributions. 

 \section{Disclosures}
 
 The institutional review board at the University of North Carolina at Chapel Hill School of Medicine approved the study and waiver of informed consent. No funding or other research support was provided by the device manufacturers. All authors take responsibility for the integrity of the data and analyses. The data that support the findings of this study are available from the corresponding author upon reasonable request.

 Dr. Rosman is the inventor of a machine learning based software for implanted cardiac devices in U.S. Provisional Patent Application 63/684,842). Dr. Rosman has received research grants from Boston Scientific and served on a Biotronik advisory board. She also reports consultancy fees from Pfizer and Biotronik.

 The content of the paper does not necessarily represent the views or policy of the Environmental Protection Agency. 

\newpage
\begin{singlespace}
 \bibliographystyle{rss}
	\bibliography{ref}

\begin{thebibliography}{59}
\expandafter\ifx\csname natexlab\endcsname\relax\def\natexlab#1{#1}\fi
\expandafter\ifx\csname url\endcsname\relax
  \def\url#1{\texttt{#1}}\fi
\expandafter\ifx\csname urlprefix\endcsname\relax\def\urlprefix{URL}\fi

\bibitem[{Ahmed et~al.(2012)Ahmed, Hossain and Doksum}]{ahmed2012lasso}
Ahmed, S.~E., Hossain, S. and Doksum, K.~A. (2012) {LASSO} and shrinkage estimation in weibull censored regression models.
\newblock \textit{Journal of Statistical Planning and Inference}, \textbf{142}, 1273--1284.

\bibitem[{Bar et~al.(2020)Bar, Booth and Wells}]{bar2020scalable}
Bar, H.~Y., Booth, J.~G. and Wells, M.~T. (2020) A scalable empirical {Bayes} approach to variable selection in generalized linear models.
\newblock \textit{Journal of Computational and Graphical Statistics}, \textbf{29}, 535--546.

\bibitem[{Bhatnagar(2017)}]{bhatnagar2017environmental}
Bhatnagar, A. (2017) Environmental determinants of cardiovascular disease.
\newblock \textit{Circulation research}, \textbf{121}, 162--180.

\bibitem[{Bhattacharya et~al.(2015)Bhattacharya, Pati, Pillai and Dunson}]{bhattacharya2015dirichlet}
Bhattacharya, A., Pati, D., Pillai, N.~S. and Dunson, D.~B. (2015) {Dirichlet--Laplace} priors for optimal shrinkage.
\newblock \textit{Journal of the American Statistical Association}, \textbf{110}, 1479--1490.

\bibitem[{Bozkurt et~al.(2024)Bozkurt, Ahmad, Alexander, Baker, Bosak, Breathett, Carter, Drazner, Dunlay, Fonarow et~al.}]{bozkurt2024hf}
Bozkurt, B., Ahmad, T., Alexander, K., Baker, W.~L., Bosak, K., Breathett, K., Carter, S., Drazner, M.~H., Dunlay, S.~M., Fonarow, G.~C. et~al. (2024) {HF STATS 2024: Heart Failure Epidemiology and Outcomes Statistics An Updated 2024 Report from the Heart Failure Society of America}.
\newblock \textit{Journal of cardiac failure}, S1071--9164.

\bibitem[{Bu{\v{c}}ar et~al.(2004)Bu{\v{c}}ar, Nagode and Fajdiga}]{buvcar2004reliability}
Bu{\v{c}}ar, T., Nagode, M. and Fajdiga, M. (2004) Reliability approximation using finite {Weibull} mixture distributions.
\newblock \textit{Reliability Engineering \& System Safety}, \textbf{84}, 241--251.

\bibitem[{Carvalho et~al.(2009)Carvalho, Polson and Scott}]{carvalho2009handling}
Carvalho, C.~M., Polson, N.~G. and Scott, J.~G. (2009) Handling sparsity via the horseshoe.
\newblock In \textit{Artificial Intelligence and Statistics}, 73--80. PMLR.

\bibitem[{Chen et~al.(2024)Chen, Dazard, Khalifa, Motairek, Kreatsoulas, Rajagopalan and Al-Kindi}]{chen2024deep}
Chen, Z., Dazard, J.-E., Khalifa, Y., Motairek, I., Kreatsoulas, C., Rajagopalan, S. and Al-Kindi, S. (2024) {Deep Learning--Based Assessment of Built Environment From Satellite Images and Cardiometabolic Disease Prevalence}.
\newblock \textit{JAMA Cardiology}.

\bibitem[{Cleland et~al.(2023)Cleland, Rosman, Hill, Mazzella, Ward-Caviness and Rappold}]{cleland2023impact}
Cleland, S.~E., Rosman, L., Hill, K.~L., Mazzella, A.~J., Ward-Caviness, C. and Rappold, A.~G. (2023) {The Impact of Temperature and Relative Humidity on Ventricular Arrhythmias in Patients with Implanted Cardiac Devices in North Carolina, 2010-2021}.
\newblock In \textit{ISEE Conference Abstracts}, vol. 2023.

\bibitem[{Dadvand et~al.(2016)Dadvand, Bartoll, Basaga{\~n}a, Dalmau-Bueno, Martinez, Ambros, Cirach, Triguero-Mas, Gascon, Borrell et~al.}]{dadvand2016green}
Dadvand, P., Bartoll, X., Basaga{\~n}a, X., Dalmau-Bueno, A., Martinez, D., Ambros, A., Cirach, M., Triguero-Mas, M., Gascon, M., Borrell, C. et~al. (2016) Green spaces and general health: roles of mental health status, social support, and physical activity.
\newblock \textit{Environment international}, \textbf{91}, 161--167.

\bibitem[{Damlen et~al.(1999)Damlen, Wakefield and Walker}]{damlen1999gibbs}
Damlen, P., Wakefield, J. and Walker, S. (1999) Gibbs sampling for {Bayesian} non-conjugate and hierarchical models by using auxiliary variables.
\newblock \textit{Journal of the Royal Statistical Society: Series B (Statistical Methodology)}, \textbf{61}, 331--344.

\bibitem[{Diebolt and Robert(1994)}]{diebolt1994estimation}
Diebolt, J. and Robert, C.~P. (1994) Estimation of finite mixture distributions through {Bayesian} sampling.
\newblock \textit{Journal of the Royal Statistical Society: Series B (Methodological)}, \textbf{56}, 363--375.

\bibitem[{Gelman et~al.(2019)Gelman, Goodrich, Gabry and Vehtari}]{gelman2019r}
Gelman, A., Goodrich, B., Gabry, J. and Vehtari, A. (2019) R-squared for {Bayesian} regression models.
\newblock \textit{The American Statistician}.

\bibitem[{Goeman(2010)}]{goeman2010l1}
Goeman, J.~J. (2010) L1 penalized estimation in the {Cox} proportional hazards model.
\newblock \textit{Biometrical journal}, \textbf{52}, 70--84.

\bibitem[{Gui and Li(2005)}]{gui2005penalized}
Gui, J. and Li, H. (2005) Penalized {Cox} regression analysis in the high-dimensional and low-sample size settings, with applications to microarray gene expression data.
\newblock \textit{Bioinformatics}, \textbf{21}, 3001--3008.

\bibitem[{Guo et~al.(2021)Guo, Wu and Cheng}]{guo2021comprehensive}
Guo, C.-Y., Wu, M.-Y. and Cheng, H.-M. (2021) The comprehensive machine learning analytics for heart failure.
\newblock \textit{International Journal of Environmental Research and Public Health}, \textbf{18}, 4943.

\bibitem[{Hastie et~al.(2009)Hastie, Tibshirani, Friedman and Friedman}]{hastie2009elements}
Hastie, T., Tibshirani, R., Friedman, J.~H. and Friedman, J.~H. (2009) \textit{The elements of statistical learning: data mining, inference, and prediction}, vol.~2.
\newblock Springer.

\bibitem[{Hoerl and Kennard(1970)}]{hoerl1970ridge}
Hoerl, A.~E. and Kennard, R.~W. (1970) Ridge regression: {Biased} estimation for nonorthogonal problems.
\newblock \textit{Technometrics}, \textbf{12}, 55--67.

\bibitem[{James et~al.(2015)James, Banay, Hart and Laden}]{james2015review}
James, P., Banay, R.~F., Hart, J.~E. and Laden, F. (2015) A review of the health benefits of greenness.
\newblock \textit{Current epidemiology reports}, \textbf{2}, 131--142.

\bibitem[{Jia et~al.(2023)Jia, Lin, He, Li, Zhang, Wang, Liu, Li, Huang, Cao et~al.}]{jia2023effect}
Jia, Y., Lin, Z., He, Z., Li, C., Zhang, Y., Wang, J., Liu, F., Li, J., Huang, K., Cao, J. et~al. (2023) Effect of air pollution on heart failure: systematic review and meta-analysis.
\newblock \textit{Environmental Health Perspectives}, \textbf{131}, 076001.

\bibitem[{Lawson et~al.(2020)Lawson, Zaccardi, Squire, Okhai, Davies, Huang, Mamas, Lam, Khunti and Kadam}]{lawson2020risk}
Lawson, C.~A., Zaccardi, F., Squire, I., Okhai, H., Davies, M., Huang, W., Mamas, M., Lam, C.~S., Khunti, K. and Kadam, U.~T. (2020) Risk factors for heart failure: 20-year population-based trends by sex, socioeconomic status, and ethnicity.
\newblock \textit{Circulation: Heart Failure}, \textbf{13}, e006472.

\bibitem[{Lee et~al.(2011)Lee, Chakraborty and Sun}]{lee2011bayesian}
Lee, K.~H., Chakraborty, S. and Sun, J. (2011) Bayesian variable selection in semiparametric proportional hazards model for high dimensional survival data.
\newblock \textit{The International Journal of Biostatistics}, \textbf{7}, 0000102202155746791301.

\bibitem[{Lee et~al.(2017)Lee, Chakraborty and Sun}]{lee2017variable}
--- (2017) Variable selection for high-dimensional genomic data with censored outcomes using group {LASSO} prior.
\newblock \textit{Computational Statistics \& Data Analysis}, \textbf{112}, 1--13.

\bibitem[{Li et~al.(2007)Li, Schneider and Bennett}]{li2007estimation}
Li, Y., Schneider, J.~A. and Bennett, D.~A. (2007) Estimation of the mediation effect with a binary mediator.
\newblock \textit{Statistics in Medicine}, \textbf{26}, 3398--3414.

\bibitem[{Liang et~al.(2023)Liang, Livingstone and Griffin}]{liang2023adaptive}
Liang, X., Livingstone, S. and Griffin, J. (2023) Adaptive {MCMC} for {Bayesian} variable selection in generalised linear models and survival models.
\newblock \textit{Entropy}, \textbf{25}, 1310.

\bibitem[{Luo et~al.(2020)Luo, Fa, Yan, Wang, Zhou, Zhang and Yu}]{luo2020high}
Luo, C., Fa, B., Yan, Y., Wang, Y., Zhou, Y., Zhang, Y. and Yu, Z. (2020) High-dimensional mediation analysis in survival models.
\newblock \textit{PLoS Computational Biology}, \textbf{16}, e1007768.

\bibitem[{Malambo et~al.(2016)Malambo, Kengne, De~Villiers, Lambert and Puoane}]{malambo2016built}
Malambo, P., Kengne, A.~P., De~Villiers, A., Lambert, E.~V. and Puoane, T. (2016) Built environment, selected risk factors and major cardiovascular disease outcomes: {A} systematic review.
\newblock \textit{PloS one}, \textbf{11}, e0166846.

\bibitem[{Mar{\'\i}n et~al.(2005)Mar{\'\i}n, Rodriguez-Bernal and Wiper}]{marin2005using}
Mar{\'\i}n, J.~M., Rodriguez-Bernal, M. and Wiper, M.~P. (2005) Using {Weibull} mixture distributions to model heterogeneous survival data.
\newblock \textit{Communications in Statistics-Simulation and Computation}, \textbf{34}, 673--684.

\bibitem[{Mazzella et~al.(2022)Mazzella, Gehi, Lampert, Buck and Rosman}]{mazzella2022effects}
Mazzella, A.~J., Gehi, A.~K., Lampert, R., Buck, S. and Rosman, L. (2022) {Effects of COVID-19 pandemic on physical activity in children and young adults with implanted devices}.
\newblock \textit{Heart Rhythm}, \textbf{19}, 165.

\bibitem[{Miao et~al.(2018)Miao, Cai, Zhang, Fan and Li}]{miao2018predictive}
Miao, F., Cai, Y.-P., Zhang, Y.-X., Fan, X.-M. and Li, Y. (2018) Predictive modeling of hospital mortality for patients with heart failure by using an improved random survival forest.
\newblock \textit{Ieee Access}, \textbf{6}, 7244--7253.

\bibitem[{Mu et~al.(2021)Mu, Liu, Kuo and Hu}]{mu2021bayesian}
Mu, J., Liu, Q., Kuo, L. and Hu, G. (2021) Bayesian variable selection for the {Cox} regression model with spatially varying coefficients with applications to {Louisiana} respiratory cancer data.
\newblock \textit{Biometrical Journal}, \textbf{63}, 1607--1622.

\bibitem[{Newcombe et~al.(2017)Newcombe, Raza~Ali, Blows, Provenzano, Pharoah, Caldas and Richardson}]{newcombe2017weibull}
Newcombe, P.~J., Raza~Ali, H., Blows, F.~M., Provenzano, E., Pharoah, P.~D., Caldas, C. and Richardson, S. (2017) Weibull regression with {Bayesian} variable selection to identify prognostic tumour markers of breast cancer survival.
\newblock \textit{Statistical Methods in Medical Research}, \textbf{26}, 414--436.

\bibitem[{Nie and Ro{\v{c}}kov{\'a}(2023)}]{nie2023bayesian}
Nie, L. and Ro{\v{c}}kov{\'a}, V. (2023) Bayesian bootstrap spike-and-slab {LASSO}.
\newblock \textit{Journal of the American Statistical Association}, \textbf{118}, 2013--2028.

\bibitem[{Pencina and D'agostino(2004)}]{pencina2004overall}
Pencina, M.~J. and D'agostino, R.~B. (2004) Overall {C} as a measure of discrimination in survival analysis: model specific population value and confidence interval estimation.
\newblock \textit{Statistics in medicine}, \textbf{23}, 2109--2123.

\bibitem[{Ro{\v{c}}kov{\'a} and George(2018)}]{rovckova2018spike}
Ro{\v{c}}kov{\'a}, V. and George, E.~I. (2018) The spike-and-slab {LASSO}.
\newblock \textit{Journal of the American Statistical Association}, \textbf{113}, 431--444.

\bibitem[{Rockova et~al.(2012)Rockova, Lesaffre, Luime and L{\"o}wenberg}]{rockova2012hierarchical}
Rockova, V., Lesaffre, E., Luime, J. and L{\"o}wenberg, B. (2012) Hierarchical {Bayesian} formulations for selecting variables in regression models.
\newblock \textit{Statistics in medicine}, \textbf{31}, 1221--1237.

\bibitem[{Roger(2013)}]{roger2013epidemiology}
Roger, V.~L. (2013) Epidemiology of heart failure.
\newblock \textit{Circulation research}, \textbf{113}, 646--659.

\bibitem[{Rosman et~al.(2024{\natexlab{a}})Rosman, Lampert, Zhuo, Li, Varma, Burg, Gaffey, Armbruster and Gehi}]{rosman2024wearable}
Rosman, L., Lampert, R., Zhuo, S., Li, Q., Varma, N., Burg, M., Gaffey, A.~E., Armbruster, T. and Gehi, A. (2024{\natexlab{a}}) {Wearable Devices, Health Care Use, and Psychological Well-Being in Patients With Atrial Fibrillation}.
\newblock \textit{Journal of the American Heart Association}, e033750.

\bibitem[{Rosman et~al.(2022)Rosman, Mazzella, Gehi, Liu, Li, Salmoirago-Blotcher, Lampert and Burg}]{rosman2022immediate}
Rosman, L., Mazzella, A.~J., Gehi, A., Liu, Y., Li, Q., Salmoirago-Blotcher, E., Lampert, R. and Burg, M.~M. (2022) Immediate and long-term effects of the covid-19 pandemic and lockdown on physical activity in patients with implanted cardiac devices.
\newblock \textit{Pacing and Clinical Electrophysiology}, \textbf{45}, 111--123.

\bibitem[{Rosman et~al.(2024{\natexlab{b}})Rosman, Mazzella, Gu, Vives, Lanctin, Natera, Gehi and Lampert}]{rosman2024patient}
Rosman, L., Mazzella, A.~J., Gu, X., Vives, C.~A., Lanctin, D., Natera, A.~C., Gehi, A. and Lampert, R. (2024{\natexlab{b}}) {A patient-centered intervention reduces disparities in remote monitoring in patients with implanted cardiac devices}.
\newblock \textit{Clinical Electrophysiology}, \textbf{10}, 316--330.

\bibitem[{Rosman et~al.(2021)Rosman, Salmoirago-Blotcher, Mahmood, Yang, Li, Mazzella, Lawrence~Klein, Bumgarner and Gehi}]{rosman2021arrhythmia}
Rosman, L., Salmoirago-Blotcher, E., Mahmood, R., Yang, H., Li, Q., Mazzella, A.~J., Lawrence~Klein, J., Bumgarner, J. and Gehi, A. (2021) Arrhythmia risk during the 2016 {US} presidential election: {The} cost of stressful politics.
\newblock \textit{Journal of the American Heart Association}, \textbf{10}, e020559.

\bibitem[{Scott and Berger(2010)}]{scott2010bayes}
Scott, J.~G. and Berger, J.~O. (2010) Bayes and empirical-{Bayes} multiplicity adjustment in the variable-selection problem.
\newblock \textit{The Annals of Statistics}, 2587--2619.

\bibitem[{Serang et~al.(2017)Serang, Jacobucci, Brimhall and Grimm}]{serang2017exploratory}
Serang, S., Jacobucci, R., Brimhall, K.~C. and Grimm, K.~J. (2017) Exploratory mediation analysis via regularization.
\newblock \textit{Structural Equation Modeling: A Multidisciplinary Journal}, \textbf{24}, 733--744.

\bibitem[{Shah et~al.(2013)Shah, Langrish, Nair, McAllister, Hunter, Donaldson, Newby and Mills}]{shah2013global}
Shah, A.~S., Langrish, J.~P., Nair, H., McAllister, D.~A., Hunter, A.~L., Donaldson, K., Newby, D.~E. and Mills, N.~L. (2013) Global association of air pollution and heart failure: a systematic review and meta-analysis.
\newblock \textit{The Lancet}, \textbf{382}, 1039--1048.

\bibitem[{Tibshirani(1996)}]{tibshirani1996regression}
Tibshirani, R. (1996) Regression shrinkage and selection via the {LASSO}.
\newblock \textit{Journal of the Royal Statistical Society Series B: Statistical Methodology}, \textbf{58}, 267--288.

\bibitem[{Tibshirani(1997)}]{tibshirani1997lasso}
--- (1997) The {LASSO} method for variable selection in the cox model.
\newblock \textit{Statistics in medicine}, \textbf{16}, 385--395.

\bibitem[{Tirschwell and Longstreth~Jr(2002)}]{tirschwell2002validating}
Tirschwell, D.~L. and Longstreth~Jr, W. (2002) Validating administrative data in stroke research.
\newblock \textit{Stroke}, \textbf{33}, 2465--2470.

\bibitem[{Van~Erp et~al.(2019)Van~Erp, Oberski and Mulder}]{van2019shrinkage}
Van~Erp, S., Oberski, D.~L. and Mulder, J. (2019) Shrinkage priors for {Bayesian} penalized regression.
\newblock \textit{Journal of Mathematical Psychology}, \textbf{89}, 31--50.

\bibitem[{VanderWeele and Vansteelandt(2014)}]{vanderweele2014mediation}
VanderWeele, T. and Vansteelandt, S. (2014) Mediation analysis with multiple mediators.
\newblock \textit{Epidemiologic methods}, \textbf{2}, 95--115.

\bibitem[{VanderWeele(2011)}]{vanderweele2011causal}
VanderWeele, T.~J. (2011) Causal mediation analysis with survival data.
\newblock \textit{Epidemiology}, \textbf{22}, 582--585.

\bibitem[{VanderWeele(2013)}]{vanderweele2013policy}
--- (2013) {Policy-relevant proportions for direct effects}.
\newblock \textit{Epidemiology}, \textbf{24}, 175--176.

\bibitem[{VanderWeele(2016)}]{vanderweele2016mediation}
--- (2016) Mediation analysis: a practitioner's guide.
\newblock \textit{Annual review of public health}, \textbf{37}, 17--32.

\bibitem[{Verweij and Van~Houwelingen(1994)}]{verweij1994penalized}
Verweij, P.~J. and Van~Houwelingen, H.~C. (1994) Penalized likelihood in {Cox} regression.
\newblock \textit{Statistics in medicine}, \textbf{13}, 2427--2436.

\bibitem[{Wang et~al.(2021)Wang, Li, Chen, Yu, Su, Hu, Liu, Wu, Yan and Su}]{wang2021machine}
Wang, Q., Li, B., Chen, K., Yu, F., Su, H., Hu, K., Liu, Z., Wu, G., Yan, J. and Su, G. (2021) Machine learning-based risk prediction of malignant arrhythmia in hospitalized patients with heart failure.
\newblock \textit{ESC Heart Failure}, \textbf{8}, 5363--5371.

\bibitem[{Yanchenko et~al.(2023)Yanchenko, Bondell and Reich}]{yanchenko2023spatial}
Yanchenko, E., Bondell, H.~D. and Reich, B.~J. (2023) Spatial regression modeling via the {R2D2} framework.
\newblock \textit{Environmetrics}, e2829.

\bibitem[{Yanchenko et~al.(2024)Yanchenko, Bondell and Reich}]{yanchenko2021r2d2}
--- (2024) The {R2D2} prior for generalized linear mixed models.
\newblock \textit{The American Statistician}, 1--20.

\bibitem[{Yuan and MacKinnon(2009)}]{yuan2009bayesian}
Yuan, Y. and MacKinnon, D.~P. (2009) Bayesian mediation analysis.
\newblock \textit{Psychological methods}, \textbf{14}, 301.

\bibitem[{Zhang et~al.(2022)Zhang, Naughton, Bondell and Reich}]{zhang2022bayesian}
Zhang, Y.~D., Naughton, B.~P., Bondell, H.~D. and Reich, B.~J. (2022) Bayesian regression using a prior on the model fit: {The R2-D2} shrinkage prior.
\newblock \textit{Journal of the American Statistical Association}, \textbf{117}, 862--874.

\bibitem[{Zou and Hastie(2005)}]{zou2005regularization}
Zou, H. and Hastie, T. (2005) Regularization and variable selection via the elastic net.
\newblock \textit{Journal of the Royal Statistical Society Series B: Statistical Methodology}, \textbf{67}, 301--320.

\end{thebibliography}
\end{singlespace}

%\printbibliography

\newpage

\appendix
\section{Appendix}

\subsection{Derivation of Weibull prior on $R^2$}

\noindent Assume that for $i \in \{1,...,n\}$, responses
    $$Y_i|\eta_i, \theta \sim Weibull(\theta, e^{\eta_i}), $$ 

\noindent where
    $$\eta_{i} = \beta_0 + \sum_{j=1}^p X_i\beta_j\text{, } \beta_j|\phi_j, W \stackrel{i.i.d}{\sim} N(0, \phi_jW)\text{ and } \sum_{i=1}^p \phi_j = 1.$$
    
\noindent Then, 
    $$\eta_{i} \sim N(\beta_{0}, W)\text{, } E(Y_i) = e^{\eta_i}\Gamma(1+\frac{1}{\theta})\text{ and } V(Y_i) = e^{2\eta_i}[\Gamma(1+\frac{2}{\theta}) - (\Gamma(1+\frac{1}{\theta}))^2].$$

\subsubsection{$R^2$ Derivation}

Recall the definition of Bayesian $R^2$ as

    \begin{equation} \label{eq:app_r2}
    R^2(W) = \frac{Var(\mu(\eta))}{Var(\mu(\eta)) + E(\sigma^2(\eta))}.
    \end{equation}

\noindent First deriving $V\{\mu(\eta)\}$ and $E\{\sigma^2(\eta)\}$:
    \vspace{\baselineskip}
    \begin{equation} \label{eq:var_r2}
        Var(\mu(\eta)) = Var(e^{\eta_i}\Gamma(1+\frac{1}{\theta})) = (\Gamma(1+\frac{1}{\theta}))^2\cdot (e^{W} - 1) \cdot e^{2\beta_0 + W} 
    \end{equation}

    \begin{equation} \label{eq:e_r2}
        E(\sigma^2(\eta)) = E(e^{2\eta_i}[\Gamma(1+\frac{2}{\theta}) - (\Gamma(1+\frac{1}{\theta}))^2]) = [\Gamma(1+\frac{2}{\theta}) - (\Gamma(1+\frac{1}{\theta}))^2]e^{2\beta_0 + W} e^W
    \end{equation}

\noindent Substituting (\ref{eq:var_r2}), (\ref{eq:e_r2}) into (\ref{eq:app_r2}):
    
    \begin{equation} 
        \begin{split}
       R^2 &= \frac{(\Gamma(1+\frac{1}{\theta}))^2\cdot (e^{W} - 1) \cdot e^{2\beta_0 + W}}{(\Gamma(1+\frac{1}{\theta}))^2\cdot (e^{W} - 1) \cdot e^{2\beta_0 + W} + [\Gamma(1+\frac{2}{\theta}) - (\Gamma(1+\frac{1}{\theta}))^2]e^{2\beta_0 + W} e^W} \\ &= \frac{(e^{W} - 1)}{\frac{\Gamma(1+\frac{2}{\theta})}{(\Gamma(1+\frac{1}{\theta}))^2}e^W -1}.
        \end{split}
    \end{equation}

\subsubsection{Derivation of the prior on $W$, $f_W(w)$}

\noindent Assume the prior on $R^2$, denoted $r$, is

    $$f_R(r) = \frac{1}{B(a,b)} \frac{(r - R^2_{min})^{a-1}(R^2_{max} - r)^{b-1}}{(R^2_{max} - R^2_{min})^{a+b-1}}\text{, for } R^2_{min}\leq r\leq R^2_{max}, $$

\noindent where $R^2_{min} = 0 \text{ and } R^2_{max} = \frac{1}{\frac{\Gamma(1+\frac{2}{\theta})}{(\Gamma(1+\frac{1}{\theta}))^2}}$.
\vspace{\baselineskip}

 \noindent Since we assume $R^2$ is a 1-to-1 transform of $W$, there exists some function, $g(.)$, where

    \begin{equation} \label{eq: invRW}
        R^2 = g^{-1}(W) = \frac{e^W - 1}{\frac{\Gamma(1+\frac{2}{\theta})}{(\Gamma(1+\frac{1}{\theta}))^2}e^W - 1}.
    \end{equation}

\noindent Now, let $c = \Gamma(1+\frac{2}{\theta})\text{ and } d = (\Gamma(1+\frac{1}{\theta}))^2$. Calculating the Jacobian of $g^{-1}(W)$ gives 

    \begin{equation} \label{eq: jacRW}
        |\frac{\partial}{\partial w} g^{-1}(w)| = \frac{d\cdot(d-c)e^w}{(c\cdot e^w - d)^2}.
    \end{equation}

  \noindent Finally, from (\ref{eq: invRW}) and (\ref{eq: jacRW}) and through change of variables,

    \begin{equation} 
        \begin{split}
        f_W(w) &= \frac{1}{B(a,b)}\frac{(\frac{e^w - 1}{\frac{c}{d}e^w - 1})^{a-1}((\frac{1}{\frac{c}{d}})-\frac{e^w - 1}{\frac{c}{d}e^w - 1})^{b-1}}{(\frac{1}{\frac{c}{d}})^{a+b-1}} \cdot |\frac{d\cdot(d-c)e^w}{(c\cdot e^w - d)^2}| \\ &= \frac{1}{B(a,b)}\frac{e^W c^a |d||d-c|(e^W-1)^{a-1}(-d^2 + cd)^{b-1}}{d^b(e^Wc-d)^{a+b}}.
        \end{split}
    \end{equation}

\subsection{Derivation of Posterior of $\beta$}

Let $Y_i$ denote survival response of the $i^{th}$ observation for $\{i \in \{1,...,n\}$. Let the prior of $\boldsymbol{\beta}$ be $\pi(\beta) \sim N(0, \Sigma)$ with censoring indicator variable $\delta_i = I(Y_i > C)$ for some fixed right censoring time $C$. Assume $Y_i$ is Weibull distributed with PDF
    
    $$f(Y_i|\theta, \boldsymbol{\beta}) = \theta^{\delta_i} y_i^{\delta_i(\theta-1)} e^{-\theta \delta_i X_i\boldsymbol{\beta}} e^{-y^\theta e^{-\theta X_i\boldsymbol{\beta}}}.$$

\noindent The full likelihood of $\boldsymbol{\beta}$ is:

    \begin{equation}
        \begin{split}
            f(\boldsymbol{\beta}|\theta) &\propto \prod_{i=1}^n \{e^{-\theta \delta_i X_i\boldsymbol{\beta}} e^{-y_i^\theta e^{-\theta X_i\boldsymbol{\beta}}}\} \cdot e^{-\frac{1}{2}\boldsymbol{\beta}^T\Sigma^{-1}\boldsymbol{\beta}} \\ &= \prod_{i=1}^ n \{e^{-y_i^\theta e^{-\theta X_i\boldsymbol{\beta}}}\} \cdot e^{-\frac{1}{2}\boldsymbol{\beta}^T\Sigma^{-1}\boldsymbol{\beta} -\theta \sum_{i=1}^n \delta_i X_i\boldsymbol{\beta}}.
        \end{split}
    \end{equation}

\subsubsection{Introducing Latent $\bold{U}$}

Introducing an auxiliary variable $\bold{U}$ gives joint PDF:

    \begin{equation}
        f(\boldsymbol{\beta}, \boldsymbol{U}|\theta) \propto \prod_{i=1}^n \{e^{-u_i} I(u_i > y_i^\theta e^{-\theta X_i\boldsymbol{\beta}})\}\cdot e^{-\frac{1}{2}\boldsymbol{\beta}^T\Sigma^{-1}\boldsymbol{\beta} -\theta \sum_{i=1}^n \delta_i X_i\boldsymbol{\beta}}.
    \end{equation}

\noindent Now, for the $i^{th}$ term, the original PDF is: 
    \begin{equation} \label{eq: origdfapp}
        f(\boldsymbol{\beta}|\theta) \propto  \{e^{- e^{\theta X_i\boldsymbol{\beta}} y_i^{\theta}}\} \cdot e^{-\frac{1}{2}\boldsymbol{\beta}^T\Sigma^{-1}\boldsymbol{\beta} -\theta X_i\beta}.
    \end{equation}
    
\noindent Adding in the latent $u_i$, this becomes
    \begin{equation}
        f(\boldsymbol{\beta}, u_i|\theta) \propto  \{e^{-u_i} I(u_i > y_i^\theta e^{-\theta X_i\boldsymbol{\beta}})\}\cdot e^{-\frac{1}{2}\boldsymbol{\beta}^T\Sigma^{-1}\boldsymbol{\beta} -\theta X_i\boldsymbol{\beta}}.
    \end{equation}

\noindent Finally, after marginalizing $u_i$:

    \begin{equation}
        \int f(\boldsymbol{\beta},u_i|\theta) \,du_i = e^{-\frac{1}{2}\boldsymbol{\beta}^T\Sigma^{-1}\boldsymbol{\beta}-\theta X_i\boldsymbol{\beta}} \int \{e^{-u_i} I(u_i > y_i^\theta e^{-\theta X_i\boldsymbol{\beta}})\}\, du_i = e^{-y_i^\theta e^{-\theta X_i\boldsymbol{\beta}}} \cdot e^{-\frac{1}{2}\boldsymbol{\beta}^T\Sigma^{-1}\boldsymbol{\beta} -\theta X_i\boldsymbol{\beta}},
    \end{equation}

\noindent we return to the original PDF (\ref{eq: origdfapp}) as desired.
    
    \vspace{\baselineskip}

\subsubsection{Calculating the full conditional of $\beta_j$}

For $j \in \{1,...,p\}$, the full conditional of $\beta_j$ is

    $$f(\beta_j|\boldsymbol{Y}, \boldsymbol{U}, \boldsymbol{\beta}_{(j)}, \Sigma) \propto \prod_{i=1}^n \{I(u_i > y_i^\theta e^{-\theta x_{i,j}\beta_j})\}\cdot e^{-\frac{1}{2\sigma_j^2}\beta_j^2 -\theta \sum_{i=1}^n \delta_i x_{i,j}\beta_j}.$$

To show this is a conjugate truncated Normal distribution, we will now break this into two parts. The first part is calculating the truncation of the distribution. Focusing on the indicator in
    $$\prod_{i=1}^n \{I(u_i > y_i^\theta e^{-\theta X_{i,j}\beta_j})\},$$

\noindent we first take the logarithm on both sides of the inequality:
$$ \log(u_i) > \log(y_i^\theta) -\theta  X_{i,j}\beta_j -\theta \sum_{l\neq j} X_{i,l}\beta_l, $$

\noindent which gives
    $$ \log(u_i) - \theta \log(y_i) + \theta \sum_{l\neq j}X_{i,l}\beta_l > -\theta X_{i,j} \beta_j.$$

    $$
    \begin{cases}
    \beta_j > -\frac{\log(u_i) - \theta \log(y_i) + \theta \sum_{l\neq j}X_{i,l}\beta_l}{ \theta X_{i,j}} & X_{i,j} > 0\\
    \beta_j < -\frac{\log(u_i) - \theta \log(y_i) + \theta \sum_{l\neq j}X_{i,l}\beta_l}{\theta X_{i,j}} & X_{i,j} < 0\\
    \text{Undefined} &  X_{i,j} = 0.
    \end{cases}
    $$

Now, the second part will focus on calculation of the mean and variance. The remainder of the full conditional is

    $$e^{-\frac{1}{2\sigma_j^2}\beta_j^2 -\theta \sum_{i=1}^n \delta_i 
 X_{i,j}\beta_j} = e^{-\frac{1}{2\sigma_j^2}[\beta_j^2 - 2\sigma^2_j \theta \sum_{i=1}^n \delta_i  X_{i,j}\beta_j]} $$
 
    $$= e^{-\frac{1}{2\sigma_j^2}[\beta_j^2 - 2\sigma^2_j \theta \sum_{i=1}^n \delta_i  X_{i,j}\beta_j + \sigma^4_j \theta^2  (\sum_{i=1}^n \delta_i  x_{i,j})^2] + \frac{\sigma_j^4 \theta^2 (\sum_{i=1}^n \delta_i  X_{i,j})^2}{2}}$$

    $$\propto e^{-\frac{1}{2\sigma_j^2}(\beta_j - \sigma^2_j \theta \sum_{i=1}^n \delta_i X_{i,j})^2} \sim N(\theta 
     \sigma^2_j \sum_{i=1}^n \delta_i  X_{i,j}, \sigma^2_j) = N(-\theta \sigma^2_j \Tilde{n} \bar{X}^*_j, \sigma^2_j).$$

    $\Tilde{n}: \text{number of uncensored observations}$

    $\bar{X}^*_j: \text{mean of uncensored observations for the $j^{th}$ covariate}$

\subsection{Trace plots of R2D2 on CHF dataset}

Convergence was achieved in the CDSR dataset for both significant and non-significant risk factors during mediator selection. We show a few relevant examples below.
\begin{figure}[ht!]
    \centering
    \includegraphics[width=1\linewidth, scale = 1.0]{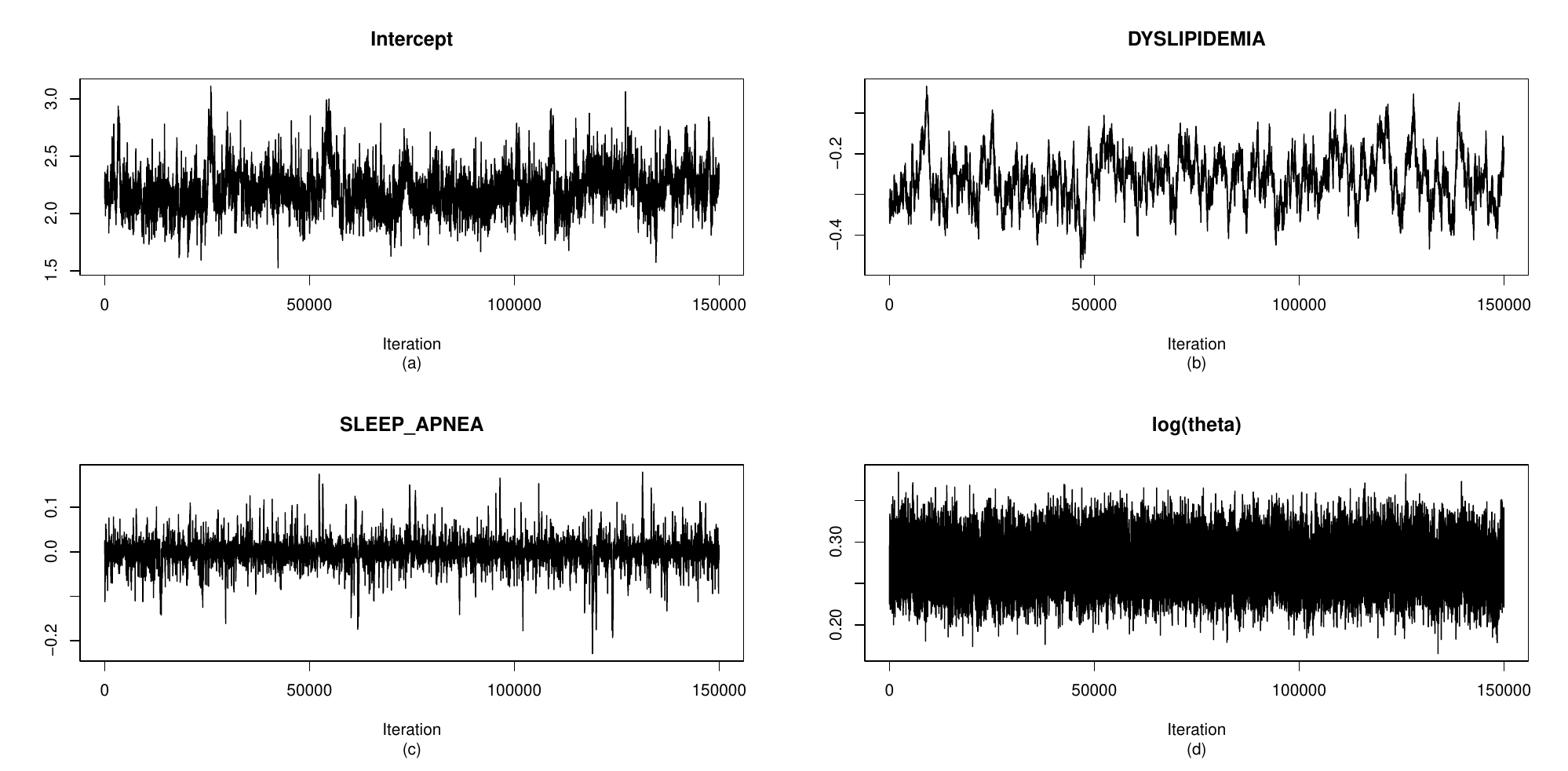}
    \caption{Trace plots of (a) the intercept, (b) a significant mediator, (c) a non-significant mediator and (d) the log of the shape parameter.}
    \label{fig:R2D2 TrPlot}
 \end{figure}

\subsection{Indirect Effect Derivation} \label{Appendix: IndEffect}

 Recall, for patient $i \in \{1,...,n\}$, let $Y_i$ represent time to CHF, $M_{j,i}$ represents the $j^{th}$ indicator for $j \in \{1,...,p_M\}$ where $p_M$ is the number of selected individual-level, and $X^*_{k,i}$ represents the community context projection for $k \in \{1,...,p_X\}$. Also recall that $\bX_{i}^* = \sum_{\ell=1}^pX_{i\ell} w_{\ell k} = \bX_i\bW$ where $p$ is the amount of community context covariates.

 The overall Weibull outcome model to estimate direct effects is:
$$ Y_i|\theta_0, \xi_0, \boldsymbol{\beta}, \boldsymbol{\tau}^D \sim \text{Weibull}\{\theta_0, \exp(\xi_0 + \sum_{j=1}^{p_M} \beta_j M_{j,i} + \sum_{k=1}^{p_X} \tau^D_k X^*_{k,i})\}.$$

\noindent The accelerated failure time parameterization allows for regression form:
$$\log(Y_i) = \sum_{j=1}^{p_M}\beta_j M_{ij} + \sum_{k=1}^{p_X}\tau_k^DX_{ik}^* = \bM_i\boldsymbol{\beta} + \bX_{i}^*\boldsymbol{\tau}^D + \epsilon_i$$

\noindent with error term $\epsilon_i$ following the GEV distribution. From \cite{li2007estimation}, for binary mediators, the indirect effect on observation $i$ is defined as 

$$\beta_j \frac{\partial E(M_j|\bX_i^*)}{\partial \bX_i^*}.$$

As we are interested in the indirect effect of the original community context covariates, we can take the partial derivative w.r.t $\bX_{i}$. Writing $E(M_j|\bX_i^*)$ in matrix form and substituting the original covariates, we have 

$$E(M_j|\bX_i^*) = \frac{\exp(\xi_j +  \bX_i^* \boldsymbol{\alpha}_j)}{1+\exp(\xi_j + \bX_i^* \boldsymbol{\alpha}_j)} = \frac{\exp(\xi_j +  \bX_i\bW \boldsymbol{\alpha}_j)}{1+\exp(\xi_j + \bX_i\bW \boldsymbol{\alpha}_j)} .$$

\noindent Now taking the derivative, 

\begin{equation} \label{eq: CC_Bin_IE}
    \frac{\partial E(M_j|\bX_i)}{\partial \bX_i} = \bW\boldsymbol{\alpha}_j  \frac{\exp(\xi_j +  \bX_i\bW \boldsymbol{\alpha}_j)}{(1+\exp(\xi_j + \bX_i\bW \boldsymbol{\alpha}_j))^2}.
\end{equation}

For continuous mediators, the indirect effect of the projections $\bX_i^*$ is $\beta_j \boldsymbol{\alpha}_j$. Transforming to the original community context covariates, $\bX_i$, this is 
\begin{equation} \label{eq: CC_Cont_IE}
    \beta_j \bW \boldsymbol{\alpha_j}.
\end{equation}

From \ref{eq: CC_Bin_IE} and \ref{eq: CC_Cont_IE}, the indirect effect of community context covariates for the $j^{th}$ mediator is therefore

\begin{equation} \label{eq: CC_IE_j}
\begin{split} 
    \boldsymbol{\tau}_j^I &= \beta_j \left[\bW \boldsymbol{\alpha_j} I(M_j \text{ continuous}) + \bW \boldsymbol{\alpha_j} \biggl\{ \frac{1}{n} \sum_{i=1}^n \frac{\exp(\xi_j +  \bX_i\bW \boldsymbol{\alpha}_j)}{(1+\exp(\xi_j + \bX_i\bW \boldsymbol{\alpha}_j))^2} \biggl\}I(M_j \text{ binary}) \right] \\
    & = \bW  \left[\beta_j \boldsymbol{\alpha_j} I(M_j \text{ cont.}) +\beta_j \boldsymbol{\alpha_j} \biggl\{ \frac{1}{n} \sum_{i=1}^n \frac{\exp(\xi_j +  \bX_i^*\boldsymbol{\alpha}_j)}{(1+\exp(\xi_j + \bX_i^* \boldsymbol{\alpha}_j))^2} \biggl\}I(M_j \text{ bin.}) \right].
\end{split}
\end{equation}

 The total indirect effect, $\boldsymbol{\omega}$, needs to sum over all $p_M$ mediators. Recall, from \ref{eq: IE_k}, the definition of $\boldsymbol{\tau}^I$ as the vector of indirect effects corresponding to all $k$ projections. Finally, given \ref{eq: CC_IE_j}, 
\begin{equation}
\begin{split}
    \boldsymbol{\omega} &=  \sum_{j=1}^{p_M} \boldsymbol{\tau}_j^I\\
    &= \sum_{j=1}^{p_M}\bW  \left[\beta_j \boldsymbol{\alpha_j} I(M_j \text{ cont.}) +\beta_j \boldsymbol{\alpha_j} \biggl\{ \frac{1}{n} \sum_{i=1}^n \frac{\exp(\xi_j +  \bX_i^*\boldsymbol{\alpha}_j)}{(1+\exp(\xi_j + \bX_i^* \boldsymbol{\alpha}_j))^2} \biggl\}I(M_j \text{ bin.}) \right]\\
    &= \bW \left[\sum_{j=1}^{p_M}\beta_j \boldsymbol{\alpha_j} I(M_j \text{ cont.}) +\beta_j \boldsymbol{\alpha_j} \biggl\{ \frac{1}{n} \sum_{i=1}^n \frac{\exp(\xi_j +  \bX_i^*\boldsymbol{\alpha}_j)}{(1+\exp(\xi_j + \bX_i^* \boldsymbol{\alpha}_j))^2} \biggl\}I(M_j \text{ bin.}) \right] \\
    &= \bW \boldsymbol{\tau}^I.
\end{split}
\end{equation}

\subsection{Selected mediator effect sizes} \label{Appendix: MedEffect}

\begin{table}[H]
\small
    \centering
    \caption{Posterior medians and 95\% Credible Intervals of the R2D2 selected covariates.}
    \label{tab: SigCov}%
    \resizebox{10cm}{!}{\begin{tabular}{ |l|c|c|}
        \hline
        Covariate & Estimate \\ 
        \hline
        Diuretics & -0.57 (-0.68, -0.44)\\
        Dyslipidemia & -0.26 (-0.38, -0.14) \\
        Atrial fibrillation \& flutter & -0.26 (-0.44, 0.00)\\
        Anticoagulants & -0.20 (-0.31, -0.08) \\
        Calcium channel blockers & -0.16 (-0.26, -0.05)\\
        Mean daily physical activity (hours) & 0.07 (0.04, 0.10) \\
        Hypertension & 0.48 (0.29, 0.67)\\
       \hline
    \end{tabular}}
\end{table}

\pagebreak

\begin{center}
\textbf{\LARGE Supplementary Materials: Mediation analysis of community context effects on heart failure using the survival R2D2 prior}
\end{center}

\section{Mixture Prior}

We derived the R2D2 prior for a Weibull mixture model on the intercept term to provide a more flexible fit in cases where the model is misspecified. The finite Weibull mixture model has been shown to be able to approximate a wide variety of distributions with positive support \citep{buvcar2004reliability}. Following \citep{marin2005using, diebolt1994estimation}, assuming we have $m$ mixture components, we redefine distribution for survival time $Y_i$ as $f_Y(y_i) \sim \sum_{j=1}^m \rho_j Weibull(\theta, e^{\eta_{ij}})$ where $j$ is the $j^{th}$ mixture component. We define $\eta_{ij} = \beta_{0j} + \sum_{k=1}^p X_i\beta_k$. This allows for $m$ intercept components where each intercept $\beta_{0j} = \beta_0 + \gamma_j$ and $\gamma_j \sim N(0, \sigma^2_\gamma)$. In the derivation of our prior, we assume $\eta_{ij} \sim N(\beta_{0j}, W)$. 

Let $h = \sum_{j=1}^m \frac{1}{m^2} [\Gamma(1 + \frac{1}{\theta})]^2e^{2\beta_{0j}} $, $k = \sum_{j=1}^m [\Gamma(1+\frac{1}{\theta})]^2[\frac{2}{m} - \frac{2}{m^2}]e^{2\beta_{0j}} $, $l =  \sum_{j=1}^m [\frac{2}{m} - \frac{1}{m^2}] [\Gamma(1+\frac{2}{\theta}) - (\Gamma(1+\frac{1}{\theta}))^2]e^{2\beta_{0j}}$. The $R^2$ is defined as 

\begin{equation} \label{eq:prior_R2mix}   
R^2 = \frac{h e^W - h}{(h + k + l)e^W - h}
\end{equation}

\noindent $R^2$ is once again bounded below by $R^2_{min} = 0$ and above by $ R^2_{max} = \frac{h}{h+k+l}$. $\Tilde{R}^2 \sim \text{Beta}(a,b,0,R^2_{max})$ is induced by

\begin{equation} \label{eq:prior_R2mix}   
\pi(w) =  \frac{1}{B(a,b)}\frac{|hke^W + hle^W|(h+k+l)^a(he^W - h)^{a-1}(hl + hk)^{b-1}}{h^{a+b-1}((h+k+l)e^W - h)^{a+b}}
\end{equation}
    
\noindent In computation, we once again substitute this exact prior for the GBP approximation. 

\subsection{Derivation of Weibull Mixture Prior on $R^2$}

    Assume there are $m$ mixture components. For $i \in \{1,...,n\}$ and $j \in \{1,...,m\}$,

    $$f_Y(y_i) \sim \sum_{j=1}^m \rho_j Weibull(\theta, e^{\eta_{ij}})$$

\noindent where

    $$\rho_1,...,\rho_m \sim Dir(\frac{1}{m},...,\frac{1}{m})\text{, } \sum_{j=1}^m \rho_j = 1$$

\noindent and

    $$\eta_{ij} = \beta_0 + \sum_{k=1}^p X_i\beta_k + \gamma_j \text{, } \beta_k|\phi_k, W \stackrel{i.i.d}{\sim} N(0, \phi_k W)\text{, } \sum_{i=1}^p \phi_k = 1.$$

\noindent Then,

    $$\eta_{ij} \sim N(\beta_{0j}, W) \text{, where } \beta_{0j} = \beta_0 + \gamma_j.$$ 
    
\subsubsection{$R^2$ Derivation}

Recall that:
    \begin{equation} \label{eq:MixR2Def}
        R^2(\boldsymbol{\beta}_0, W) = \frac{V\{\mu(\eta)\}}{V\{\mu(\eta)\} + E\{\sigma^2(\eta)\}}.
    \end{equation}

\noindent First deriving $\mu(\eta)$ and $\sigma^2(\eta)$ we have:

\begin{equation} \label{eq:MixMu}
    \begin{split}
        \mu(\eta) &= E(Y_i) = E[E(Y_i|\rho_1,...,\rho_m)] = E[\sum_{j=1}^m \rho_j E\{Weibull(\theta, e^{\eta_{ij}})\}] \\ &= E[\sum_{j=1}^m \rho_j e^{\eta_{ij}}\Gamma(1+\frac{1}{\theta})] = \sum_{j=1}^m \frac{1}{m}e^{\eta_{ij}}\Gamma(1+\frac{1}{\theta}),
    \end{split}
\end{equation}

\begin{equation} \label{eq:MixSig}
    \begin{split}
        \sigma^2(\eta) &= V(Y_i) = V[E(Y_i|\rho_1,...,\rho_m)] + E[V(Y_i|\rho_1,...,\rho_m)] \\ &= V[\sum_{j=1}^m \rho_j E\{Weibull(\theta, e^{\eta_{ij}})\}] + E[\sum_{j=1}^m \rho_j^2V\{Weibull(\theta, e^{\eta_{ij}})\}] \\ \text{(assume indep.)} &= \sum_{j=1}^m [E\{Weibull(\theta, e^{\eta_{ij}})\}]^2 V(\rho_j) + E[\sum_{j=1}^m \rho_j^2 e^{2\eta_{ij}}[\Gamma(1+\frac{2}{\theta}) - \{\Gamma(1+\frac{1}{\theta})\}^2]] \\ & = \sum_{j=1}^m e^{2\eta_{ij}}[\Gamma(1+\frac{1}{\theta})]^2[\frac{\frac{1}{m}(1-\frac{1}{m})}{2}]\\
        &+ \sum_{j=1}^m [\frac{\frac{1}{m}(1-\frac{1}{m})}{2} + \frac{1}{m^2}]e^{2\eta_{ij}}[\Gamma(1+\frac{2}{\theta}) - \{\Gamma(1+\frac{1}{\theta})\}^2]. 
    \end{split}
\end{equation}

\noindent Next, deriving $V\{\mu(\eta)\}$ and $E\{\sigma^2(\eta)\}$:

\begin{equation} \label{eq:MixVMu}
    \begin{split}
        V\{\mu(\eta)\} = V[\sum_{j=1}^m \frac{1}{m}e^{\eta_{ij}}\Gamma(1+\frac{1}{\theta})] = \sum_{j=1}^m \frac{1}{m^2} [\Gamma(1 + \frac{1}{\theta_j})]^2 (e^W-1)e^{2\beta_{0j} + W}.
    \end{split}
\end{equation}

\begin{equation} \label{eq:MixESig}
    \begin{split}
        E\{\sigma^2(\eta)\} &= E\biggr[\sum_{j=1}^m e^{2\eta_{ij}}[\Gamma(1+\frac{1}{\theta})]^2[\frac{\frac{1}{m}(1-\frac{1}{m})}{2}] \\ &+ \sum_{j=1}^m [\frac{\frac{1}{m}(1-\frac{1}{m})}{2} + \frac{1}{m^2}]e^{2\eta_{ij}}[\Gamma(1+\frac{2}{\theta}) - \{\Gamma(1+\frac{1}{\theta})\}^2]\biggr]\\  &= \sum_{j=1}^m [\Gamma(1+\frac{1}{\theta})]^2[\frac{\frac{1}{m}(1-\frac{1}{m})}{2}] e^{2\beta_{0j} + 2W} \\ &+ \sum_{j=1}^m [\frac{\frac{1}{m}(1-\frac{1}{m})}{2} + \frac{1}{m^2}] [\Gamma(1+\frac{2}{\theta}) - \{\Gamma(1+\frac{1}{\theta})\}^2] e^{2\beta_{0j} + 2W}.
    \end{split}
\end{equation}

\noindent Now, let:

    $$h = \sum_{j=1}^m \frac{1}{m^2} [\Gamma(1 + \frac{1}{\theta})]^2e^{2\beta_{0j}}, $$
    
    $$k = \sum_{j=1}^m [\Gamma(1+\frac{1}{\theta})]^2[\frac{2}{m} - \frac{2}{m^2}]e^{2\beta_{0j}}, $$
    
    $$l =  \sum_{j=1}^m [\frac{2}{m} - \frac{1}{m^2}] [\Gamma(1+\frac{2}{\theta}) - (\Gamma(1+\frac{1}{\theta}))^2]e^{2\beta_{0j}}.$$
    
\noindent Finally, substituting (\ref{eq:MixVMu}), (\ref{eq:MixESig}) into \ref{eq:MixR2Def},

\begin{equation}
    R^2 = \frac{h (e^W-1)}{h (e^W-1) + ke^{W} + l e^{W}} = \frac{h e^W - h}{(h + k + l)e^W - h}.
\end{equation}

\subsubsection{Derivation of the prior on $W$, $f_W(w)$}

\noindent Assume the prior on $R^2$, denoted $r$, is

    $$f_R(r) = \frac{1}{B(a,b)} \frac{(r - R^2_{min})^{a-1}(R^2_{max} - r)^{b-1}}{(R^2_{max} - R^2_{min})^{a+b-1}}\text{, for } R^2_{min}\leq r\leq R^2_{max}, $$

\noindent where $R^2_{min} = 0 \text{ and } R^2_{max} = \frac{h}{h+k+l}$.

\noindent Since we assume $R^2$ is a 1-to-1 transform of $W$,

    \begin{equation} \label{eq: invMixRW}
        R^2 = g^{-1}(W) = \frac{h e^W - h}{(h + k + l)e^W - h}.
    \end{equation}

\noindent Deriving the Jacobian:
    \begin{equation} \label{eq: jacMixRW}
        |\frac{\partial}{\partial w} g^{-1}(w)| = |\frac{hke^W + hle^W}{((h+k+l)e^W-h)^2}|.
    \end{equation}

\noindent From (\ref{eq: invMixRW}) and (\ref{eq: jacMixRW}), through change of variables,

    \begin{equation} 
        \begin{split}
        f_W(w) &= \frac{1}{B(a,b)} \frac{(\frac{h e^W - h}{(h + k + l)e^W - h})^{a-1}(\frac{h}{h+k+l} - \frac{h e^W - h}{(h + k + l)e^W - h})^{b-1}}{(\frac{h}{h+k+l})^{a+b-1}} \cdot |\frac{hke^W + hle^W}{((h+k+l)e^W-h)^2}| \\ &= \frac{1}{B(a,b)}\frac{|hke^W + hle^W|(h+k+l)^a(he^W - h)^{a-1}(hl + hk)^{b-1}}{h^{a+b-1}((h+k+l)e^W - h)^{a+b}}.
        \end{split}
    \end{equation}

\section{Computational Details for CHF analysis}

\subsection{R2D2 Mediator Selection}
 The dataset is first split into a 90\% training set and 10\% test set to analyze predictive performance. R2D2 is fit with $a = 0.5, b = 0.5$ on the training set for 250,000 iterations with the first 100,000 removed as burn in. 

\subsection{Mediator Models}
 For both mediation analyses, we select uninformative priors with 5,000 burn in and 20,000 MCMC iterations. For the Weibull survival model, the prior setting is 
 
    \begin{align*} 
        Y_i|\eta_i, \theta &\sim Weibull(\theta, e^{\beta_0 + X_i\boldsymbol{\beta}}) \\ 
        \beta_j &\sim N(0, 100)\text{, } j \in \{0,...,p\} \tag{\theequation}\label{eq: WeibullReg_prior}\\ 
        \log(\theta) &\sim N(0, 1000). 
    \end{align*}

 \noindent For the Logistic model, the prior setting is 
 
     \begin{align*} 
        P(Y_i = 1|\boldsymbol{\beta}, \sigma^2) &\sim \text{expit}(X_i\boldsymbol{\beta}) \\ 
        \boldsymbol{\beta}|\sigma^2 &\sim N(0, \sigma^2I_p) \tag{\theequation}\label{eq: LogReg_prior}\\ 
        \sigma^2 &\sim InvGamma(0.1,0.1). 
    \end{align*}

 \noindent For the continuous model, the prior setting is 
    \begin{align*} 
        Y_i|\boldsymbol{\beta}, \sigma^2 &\sim N(X_i\boldsymbol{\beta}, \sigma^2) \\ 
        \boldsymbol{\beta}|\tau^2 &\sim N(0, \tau^2I_p) \tag{\theequation}\label{eq: Reg_prior}\\ 
        \tau^2, \sigma^2 &\sim InvGamma(0.1,0.1). 
    \end{align*}

\section{Data Summary} \label{Appendix: Table1}

% Table 1
\begin{landscape}
\scriptsize
\begin{longtable}{l@{\hskip -1cm}llll}
\caption{Summary of CHF dataset. Column 0 represents patients who did not experience CHF while column 1 represents patients who did.} \label{tab:Combined_Data_Summary} \\
    \hline
     & 0 (N=1365) & 1 (N=1212) & Total (N=2577) & p value \\ 
    \hline
    \endfirsthead

    \caption*{\textit{Continued from previous page}} \\
    \hline
     & 0 (N=1365) & 1 (N=1212) & Total (N=2577) & p value \\ 
    \hline
    \endhead

    \hline \multicolumn{5}{r}{\textit{Continued on next page}} \\ \hline
    \endfoot

    \hline
    \endlastfoot
    \scriptsize{$^*$ This represents average years to end of study. They are censored values in the analysis.}\\
    Years to CHF &  &  &  & $<$ 0.001 \\ 
    -  Mean (SD) & 3.378 (2.334)$^*$ & 2.901 (2.256) & 3.154 (2.309) &  \\ 
    -  Range & 0.003 - 11.266 & 0.005 - 11.085 & 0.003 - 11.266 &  \\ 
    Age &  &  &  & $<$ 0.001 \\ 
    -  Mean (SD) & 76.685 (12.489) & 80.483 (10.740) & 78.471 (11.850) &  \\ 
    -  Range & 18.000 - 107.000 & 18.000 - 104.000 & 18.000 - 107.000 &  \\ 
    Sex (Male = 1) &  &  &  & 0.057 \\ 
    -  0 & 665 (48.7\%) & 636 (52.5\%) & 1301 (50.5\%) &  \\ 
    -  1 & 700 (51.3\%) & 576 (47.5\%) & 1276 (49.5\%) &  \\ 
    Race: Asian &  &  &  & 0.109 \\ 
    -  0 & 1349 (98.8\%) & 1205 (99.4\%) & 2554 (99.1\%) &  \\ 
    -  1 & 16 (1.2\%) & 7 (0.6\%) & 23 (0.9\%) &  \\ 
    Race: Black &  &  &  & 0.170 \\ 
    -  0 & 1175 (86.1\%) & 1020 (84.2\%) & 2195 (85.2\%) &  \\ 
    -  1 & 190 (13.9\%) & 192 (15.8\%) & 382 (14.8\%) &  \\ 
    Race: Other &  &  &  & 0.961 \\ 
    -  0 & 1341 (98.2\%) & 1191 (98.3\%) & 2532 (98.3\%) &  \\ 
    -  1 & 24 (1.8\%) & 21 (1.7\%) & 45 (1.7\%) &  \\ 
    Race: White &  &  &  & 0.427 \\ 
    -  0 & 237 (17.4\%) & 225 (18.6\%) & 462 (17.9\%) &  \\ 
    -  1 & 1128 (82.6\%) & 987 (81.4\%) & 2115 (82.1\%) &  \\ 
    Hispanic/Latino &  &  &  & 0.514 \\ 
    -  0 & 26 (1.9\%) & 19 (1.6\%) & 45 (1.7\%) &  \\ 
    -  1 & 1339 (98.1\%) & 1193 (98.4\%) & 2532 (98.3\%) &  \\ 
    Marital Status &  &  &  & $<$ 0.001 \\ 
    -  0 & 805 (59.0\%) & 578 (47.7\%) & 1383 (53.7\%) &  \\ 
    -  1 & 560 (41.0\%) & 634 (52.3\%) & 1194 (46.3\%) &  \\ 
    Atrial Fibrilation and Flutter &  &  &  & 0.408 \\ 
    -  0 & 1228 (90.0\%) & 1102 (90.9\%) & 2330 (90.4\%) &  \\ 
    -  1 & 137 (10.0\%) & 110 (9.1\%) & 247 (9.6\%) &  \\ 
     CABG &  &  &  & 0.551 \\ 
      -  0 & 1248 (91.4\%) & 1100 (90.8\%) & 2348 (91.1\%) &  \\ 
      -  1 & 117 (8.6\%) & 112 (9.2\%) & 229 (8.9\%) &  \\ 
      CAD &  &  &  & 0.106 \\ 
      -  0 & 1147 (84.0\%) & 1046 (86.3\%) & 2193 (85.1\%) &  \\ 
      -  1 & 218 (16.0\%) & 166 (13.7\%) & 384 (14.9\%) &  \\ 
      Cancer &  &  &  & 0.658 \\ 
      -  0 & 1284 (94.1\%) & 1145 (94.5\%) & 2429 (94.3\%) &  \\ 
      -  1 & 81 (5.9\%) & 67 (5.5\%) & 148 (5.7\%) &  \\ 
      COPD &  &  &  & 0.657 \\ 
      -  0 & 1216 (89.1\%) & 1073 (88.5\%) & 2289 (88.8\%) &  \\ 
      -  1 & 149 (10.9\%) & 139 (11.5\%) & 288 (11.2\%) &  \\ 
      Depression &  &  &  & 0.425 \\ 
      -  0 & 1281 (93.8\%) & 1128 (93.1\%) & 2409 (93.5\%) &  \\ 
      -  1 & 84 (6.2\%) & 84 (6.9\%) & 168 (6.5\%) &  \\ 
      Diabetes &  &  &  & $<$ 0.001 \\ 
      -  0 & 1209 (88.6\%) & 1139 (94.0\%) & 2348 (91.1\%) &  \\ 
      -  1 & 156 (11.4\%) & 73 (6.0\%) & 229 (8.9\%) &  \\ 
      Dyslipidemia &  &  &  & 0.670 \\ 
      -  0 & 932 (68.3\%) & 837 (69.1\%) & 1769 (68.6\%) &  \\ 
      -  1 & 433 (31.7\%) & 375 (30.9\%) & 808 (31.4\%) &  \\ 
      Hypertension &  &  &  & $<$ 0.001 \\ 
      -  0 & 1024 (75.0\%) & 1121 (92.5\%) & 2145 (83.2\%) &  \\ 
      -  1 & 341 (25.0\%) & 91 (7.5\%) & 432 (16.8\%) &  \\ 
      Myocardial Infarction &  &  &  & 0.787 \\ 
      -  0 & 1140 (83.5\%) & 1017 (83.9\%) & 2157 (83.7\%) &  \\ 
      -  1 & 225 (16.5\%) & 195 (16.1\%) & 420 (16.3\%) &  \\ 
      Nonrheumatic Valve Disorders &  &  &  & 0.497 \\ 
      -  0 & 1097 (80.4\%) & 961 (79.3\%) & 2058 (79.9\%) &  \\ 
      -  1 & 268 (19.6\%) & 251 (20.7\%) & 519 (20.1\%) &  \\ 
      Peripheral Vascular Heart Disease &  &  &  & 0.021 \\ 
      -  0 & 1193 (87.4\%) & 1021 (84.2\%) & 2214 (85.9\%) &  \\ 
      -  1 & 172 (12.6\%) & 191 (15.8\%) & 363 (14.1\%) &  \\ 
      Renal Failure/Chronic Kidney Disease &  &  &  & $<$ 0.001 \\ 
      -  0 & 1307 (95.8\%) & 1095 (90.3\%) & 2402 (93.2\%) &  \\ 
      -  1 & 58 (4.2\%) & 117 (9.7\%) & 175 (6.8\%) &  \\ 
     Sleep Apnea &  &  &  & 0.004 \\ 
      -  0 & 1272 (93.2\%) & 1091 (90.0\%) & 2363 (91.7\%) &  \\ 
      -  1 & 93 (6.8\%) & 121 (10.0\%) & 214 (8.3\%) &  \\ 
      Sudden Cardiac Arrest &  &  &  & 0.393 \\ 
      -  0 & 1344 (98.5\%) & 1188 (98.0\%) & 2532 (98.3\%) &  \\ 
      -  1 & 21 (1.5\%) & 24 (2.0\%) & 45 (1.7\%) &  \\ 
      Valvular Heart Diseases &  &  &  & $<$ 0.001 \\ 
      -  0 & 1041 (76.3\%) & 1025 (84.6\%) & 2066 (80.2\%) &  \\ 
      -  1 & 324 (23.7\%) & 187 (15.4\%) & 511 (19.8\%) &  \\ 
      Mitral Regurgitation &  &  &  & $<$ 0.001 \\ 
      -  0 & 1108 (81.2\%) & 907 (74.8\%) & 2015 (78.2\%) &  \\ 
      -  1 & 257 (18.8\%) & 305 (25.2\%) & 562 (21.8\%) &  \\ 
      Generalized Anxiety Disorder &  &  &  & 0.907 \\ 
      -  0 & 1351 (99.0\%) & 1199 (98.9\%) & 2550 (99.0\%) &  \\ 
      -  1 & 14 (1.0\%) & 13 (1.1\%) & 27 (1.0\%) &  \\ 
      Panic Disorder &  &  &  & 0.313 \\ 
      -  0 & 1359 (99.6\%) & 1203 (99.3\%) & 2562 (99.4\%) &  \\ 
      -  1 & 6 (0.4\%) & 9 (0.7\%) & 15 (0.6\%) &  \\ 
      Entresto &  &  &  & 0.006 \\ 
      -  0 & 1310 (96.0\%) & 1134 (93.6\%) & 2444 (94.8\%) &  \\ 
      -  1 & 55 (4.0\%) & 78 (6.4\%) & 133 (5.2\%) &  \\ 
      SGLT2I &  &  &  & 0.069 \\ 
      -  0 & 1343 (98.4\%) & 1180 (97.4\%) & 2523 (97.9\%) &  \\ 
      -  1 & 22 (1.6\%) & 32 (2.6\%) & 54 (2.1\%) &  \\ 
      Aspirin &  &  &  & 0.512 \\ 
      -  0 & 693 (50.8\%) & 631 (52.1\%) & 1324 (51.4\%) &  \\ 
      -  1 & 672 (49.2\%) & 581 (47.9\%) & 1253 (48.6\%) &  \\ 
      Alpha Beta Blockers &  &  &  & $<$ 0.001 \\ 
      -  0 & 1298 (95.1\%) & 1057 (87.2\%) & 2355 (91.4\%) &  \\ 
      -  1 & 67 (4.9\%) & 155 (12.8\%) & 222 (8.6\%) &  \\ 
      Beta Blockers &  &  &  & $<$ 0.001 \\ 
      -  0 & 853 (62.5\%) & 618 (51.0\%) & 1471 (57.1\%) &  \\ 
      -  1 & 512 (37.5\%) & 594 (49.0\%) & 1106 (42.9\%) &  \\
      ACE Inhibitors &  &  &  & 0.001 \\ 
      -  0 & 1016 (74.4\%) & 833 (68.7\%) & 1849 (71.8\%) &  \\ 
      -  1 & 349 (25.6\%) & 379 (31.3\%) & 728 (28.2\%) &  \\ 
      ARB &  &  &  & $<$ 0.001 \\ 
      -  0 & 1112 (81.5\%) & 910 (75.1\%) & 2022 (78.5\%) &  \\ 
      -  1 & 253 (18.5\%) & 302 (24.9\%) & 555 (21.5\%) &  \\ 
      Anti Arrhythmic &  &  &  & $<$ 0.001 \\ 
      -  0 & 1209 (88.6\%) & 975 (80.4\%) & 2184 (84.7\%) &  \\ 
      -  1 & 156 (11.4\%) & 237 (19.6\%) & 393 (15.3\%) &  \\ 
      Calcium Channel Blockers &  &  &  & $<$ 0.001 \\ 
      -  0 & 936 (68.6\%) & 720 (59.4\%) & 1656 (64.3\%) &  \\ 
      -  1 & 429 (31.4\%) & 492 (40.6\%) & 921 (35.7\%) &  \\ 
      Statins &  &  &  & $<$ 0.001 \\ 
      -  0 & 690 (50.5\%) & 527 (43.5\%) & 1217 (47.2\%) &  \\ 
      -  1 & 675 (49.5\%) & 685 (56.5\%) & 1360 (52.8\%) &  \\ 
      Platelet Inhibitors &  &  &  & 0.012 \\ 
      -  0 & 653 (47.8\%) & 520 (42.9\%) & 1173 (45.5\%) &  \\ 
      -  1 & 712 (52.2\%) & 692 (57.1\%) & 1404 (54.5\%) &  \\ 
      Anticoagulant &  &  &  & $<$ 0.001 \\ 
      -  0 & 1057 (77.4\%) & 822 (67.8\%) & 1879 (72.9\%) &  \\ 
      -  1 & 308 (22.6\%) & 390 (32.2\%) & 698 (27.1\%) &  \\ 
      Antidepressant &  &  &  & 0.077 \\ 
      -  0 & 1106 (81.0\%) & 948 (78.2\%) & 2054 (79.7\%) &  \\ 
      -  1 & 259 (19.0\%) & 264 (21.8\%) & 523 (20.3\%) &  \\ 
      Statin/CCB Combo &  &  &  & 0.237 \\ 
      -  0 & 1362 (99.8\%) & 1206 (99.5\%) & 2568 (99.7\%) &  \\ 
      -  1 & 3 (0.2\%) & 6 (0.5\%) & 9 (0.3\%) &  \\ 
     ACE Inhib/CCB Combo &  &  &  & 0.216 \\ 
      -  0 & 1352 (99.0\%) & 1194 (98.5\%) & 2546 (98.8\%) &  \\ 
      -  1 & 13 (1.0\%) & 18 (1.5\%) & 31 (1.2\%) &  \\
     ARB/CCB Combo &  &  &  & 0.367 \\ 
      -  0 & 1354 (99.2\%) & 1198 (98.8\%) & 2552 (99.0\%) &  \\ 
      -  1 & 11 (0.8\%) & 14 (1.2\%) & 25 (1.0\%) &  \\ 
      Digoxin &  &  &  & $<$ 0.001 \\ 
      -  0 & 1351 (99.0\%) & 1159 (95.6\%) & 2510 (97.4\%) &  \\ 
      -  1 & 14 (1.0\%) & 53 (4.4\%) & 67 (2.6\%) &  \\ 
      Diuretics &  &  &  & $<$ 0.001 \\ 
      -  0 & 1191 (87.3\%) & 674 (55.6\%) & 1865 (72.4\%) &  \\ 
      -  1 & 174 (12.7\%) & 538 (44.4\%) & 712 (27.6\%) &  \\ 
      BMI &  &  &  & $<$ 0.001 \\ 
      -  Mean (SD) & 28.142 (6.131) & 29.055 (7.153) & 28.572 (6.646) &  \\ 
      -  Range & 14.650 - 58.950 & 14.620 - 63.080 & 14.620 - 63.080 &  \\ 
      Former Smoker &  &  &  & $<$ 0.001 \\ 
      -  0 & 797 (58.4\%) & 609 (50.2\%) & 1406 (54.6\%) &  \\ 
      -  1 & 568 (41.6\%) & 603 (49.8\%) & 1171 (45.4\%) &  \\ 
      Never Smoker &  &  &  & $<$ 0.001 \\ 
      -  0 & 664 (48.6\%) & 675 (55.7\%) & 1339 (52.0\%) &  \\ 
      -  1 & 701 (51.4\%) & 537 (44.3\%) & 1238 (48.0\%) &  \\ 
      Mean daily physical activity (min) &  &  &  & $<$ 0.001 \\ 
      -  Mean (SD) & 2.609 (1.857) & 1.935 (1.692) & 2.292 (1.813) &  \\ 
      -  Range & 0.000 - 9.683 & 0.000 - 8.750 & 0.000 - 9.683 &  \\
      Urban Area Y/N &  &  &  & 0.608 \\ 
      -  0 & 579 (42.4\%) & 502 (41.4\%) & 1081 (41.9\%) &  \\ 
      -  1 & 786 (57.6\%) & 710 (58.6\%) & 1496 (58.1\%) &  \\ 
      Within 1km Urban Area Y/N &  &  &  & 0.969 \\ 
      -  0 & 454 (33.3\%) & 404 (33.3\%) & 858 (33.3\%) &  \\ 
      -  1 & 911 (66.7\%) & 808 (66.7\%) & 1719 (66.7\%) &  \\
      Mean \% urban imperviousness cover &  &  &  & 0.496 \\ 
      -  Mean (SD) & 13.327 (11.632) & 13.014 (11.649) & 13.180 (11.639) &  \\ 
      -  Range & 0.030 - 56.950 & 0.190 - 53.990 & 0.030 - 56.950 &  \\ 
      Mean \% tree canopy cover &  &  &  & $<$ 0.001 \\ 
      -  Mean (SD) & 42.336 (15.929) & 40.156 (15.298) & 41.311 (15.670) &  \\ 
      -  Range & 6.500 - 88.560 & 5.620 - 88.390 & 5.620 - 88.560 &  \\ 
      \% Open water \& emergent herbaceous wetlands &  &  &  & 0.059 \\ 
      -  Mean (SD) & 1.619 (4.934) & 1.314 (2.858) & 1.475 (4.093) &  \\ 
      -  Range & 0.000 - 74.620 & 0.000 - 40.440 & 0.000 - 74.620 &  \\ 
      \% Forest \& woody wetlands &  &  &  & 0.632 \\ 
      -  Mean (SD) & 32.766 (20.245) & 32.392 (19.197) & 32.590 (19.756) &  \\ 
      -  Range & 0.000 - 94.680 & 0.000 - 90.440 & 0.000 - 94.680 &  \\ 
      \% Barren land \& shrub \& grassland &  &  &  & 0.037 \\ 
      -  Mean (SD) & 3.552 (4.125) & 3.889 (4.057) & 3.711 (4.096) &  \\ 
      -  Range & 0.000 - 31.030 & 0.000 - 26.670 & 0.000 - 31.030 &  \\ 
      \% Agricultural land &  &  &  & 0.002 \\ 
      -  Mean (SD) & 16.597 (18.947) & 18.940 (18.776) & 17.699 (18.899) &  \\ 
      -  Range & 0.000 - 78.970 & 0.000 - 82.280 & 0.000 - 82.280 &  \\ 
      \% Developed open spaces \& low intensity &  &  &  & 0.051 \\ 
      -  Mean (SD) & 35.299 (23.150) & 33.548 (22.252) & 34.475 (22.745) &  \\ 
      -  Range & 0.320 - 92.180 & 2.630 - 89.310 & 0.320 - 92.180 &  \\ 
      \% Developed medium \& high intensity &  &  &  & 0.570 \\ 
      -  Mean (SD) & 10.168 (11.259) & 9.917 (11.154) & 10.050 (11.209) &  \\ 
      -  Range & 0.000 - 62.940 & 0.000 - 57.080 & 0.000 - 62.940 &  \\ 
      EJ Index for $\text{PM}_{\text{2.5}}$ level &  &  &  & 0.111 \\ 
      -  Mean (SD) & -1126.975 (4884.487) & -818.226 (4937.798) & -981.766 (4911.097) &  \\ 
      -  Range & -18575.787 - 20531.934 & -18575.787 - 20531.934 & -18575.787 - 20531.934 &  \\ 
      EJ Index for Ozone level&  &  &  & 0.113 \\ 
      -  Mean (SD) & -4541.460 (19997.539) & -3286.814 (20165.352) & -3951.382 (20082.507) &  \\ 
      -  Range & -76032.548 - 82422.312 & -76032.548 - 82422.312 & -76032.548 - 82422.312 &  \\ 
      EJ Index for Diesel level &  &  &  & 0.183 \\ 
      -  Mean (SD) & -77.914 (328.713) & -60.508 (333.023) & -69.727 (330.797) &  \\ 
      -  Range & -1198.840 - 2936.865 & -1198.840 - 2936.865 & -1198.840 - 2936.865 &  \\ 
      EJ Index for Air toxics cancer risk &  &  &  & 0.114 \\ 
      -  Mean (SD) & -4849.824 (19693.154) & -3613.843 (19900.312) & -4268.524 (19796.625) &  \\ 
      -  Range & -67793.262 - 90976.144 & -67793.262 - 90976.144 & -67793.262 - 90976.144 &  \\ 
     EJ Index for Air toxics respiratory hazard index  &  &  &  & 0.098 \\ 
      -  Mean (SD) & -159.307 (650.534) & -116.661 (655.601) & -139.250 (653.142) &  \\ 
      -  Range & -2420.890 - 3544.837 & -2420.890 - 3544.837 & -2420.890 - 3544.837 &  \\ 
      EJ Index for Traffic  &  &  &  & 0.275 \\ 
      -  Mean (SD) & -195.710 (170836.445) & 7261.131 (175311.039) & 3311.349 (172961.718) &  \\ 
      -  Range & -674536.316 - 3435968.650 & -462893.857 - 3435968.650 & -674536.316 - 3435968.650 &  \\ 
      EJ Index for Lead paint &  &  &  & 0.067 \\ 
      -  Mean (SD) & 4.567 (66.916) & 9.372 (66.136) & 6.827 (66.581) &  \\ 
      -  Range & -357.515 - 488.224 & -357.515 - 456.821 & -357.515 - 488.224 &  \\ 
      EJ Index for National Priorities List site proximity &  &  &  & 0.132 \\ 
      -  Mean (SD) & -11.087 (42.883) & -8.603 (40.425) & -9.919 (41.755) &  \\ 
      -  Range & -584.170 - 286.667 & -271.531 - 334.503 & -584.170 - 334.503 &  \\ 
      EJ Index for Risk Management Plan site proximity &  &  &  & 0.007 \\ 
      -  Mean (SD) & 8.357 (186.932) & 32.441 (259.853) & 19.684 (224.479) &  \\ 
      -  Range & -1274.749 - 1461.026 & -1274.749 - 2282.517 & -1274.749 - 2282.517 &  \\ 
      EJ Index for Treatment Storage and Disposal site proximity &  &  &  & 0.017 \\ 
      -  Mean (SD) & -33.420 (290.472) & 2.015 (452.587) & -16.755 (375.878) &  \\ 
      -  Range & -1556.501 - 3550.490 & -1548.996 - 10053.009 & -1556.501 - 10053.009 &  \\ 
      EJ Index for water discharge site proximity &  &  &  & 0.881 \\ 
      -  Mean (SD) & -0.838 (28.450) & -1.001 (26.734) & -0.914 (27.651) &  \\ 
      -  Range & -315.490 - 562.593 & -392.098 - 321.009 & -392.098 - 562.593 &  \\ 
      National Walkability Index &  &  &  & 0.220 \\ 
      -  Mean (SD) & 7.353 (3.592) & 7.182 (3.475) & 7.272 (3.537) &  \\ 
      -  Range & 1.500 - 19.667 & 1.500 - 18.500 & 1.500 - 19.667 &  \\ 
      Median home age &  &  &  & 0.096 \\ 
      -  Mean (SD) & 28.504 (10.553) & 29.196 (10.529) & 28.830 (10.546) &  \\ 
      -  Range & 9.000 - 76.000 & 9.000 - 70.000 & 9.000 - 76.000 &  \\
      \% Poverty &  &  &  & $<$ 0.001 \\ 
      -  Mean (SD) & 15.193 (9.194) & 16.573 (9.964) & 15.842 (9.587) &  \\ 
      -  Range & 0.370 - 56.485 & 0.000 - 58.897 & 0.000 - 58.897 &  \\
      Primary rural-urban commuting area code &  &  &  & 0.793 \\ 
      -  1 & 673 (49.3\%) & 556 (45.9\%) & 1229 (47.7\%) &  \\ 
      -  2 & 262 (19.2\%) & 247 (20.4\%) & 509 (19.8\%) &  \\ 
      -  3 & 22 (1.6\%) & 17 (1.4\%) & 39 (1.5\%) &  \\ 
      -  4 & 144 (10.5\%) & 144 (11.9\%) & 288 (11.2\%) &  \\ 
      -  5 & 56 (4.1\%) & 49 (4.0\%) & 105 (4.1\%) &  \\ 
      -  6 & 21 (1.5\%) & 22 (1.8\%) & 43 (1.7\%) &  \\ 
      -  7 & 88 (6.4\%) & 94 (7.8\%) & 182 (7.1\%) &  \\ 
      -  8 & 37 (2.7\%) & 30 (2.5\%) & 67 (2.6\%) &  \\ 
      -  9 & 41 (3.0\%) & 37 (3.1\%) & 78 (3.0\%) &  \\ 
      -  10 & 21 (1.5\%) & 16 (1.3\%) & 37 (1.4\%) &  \\ 
      Secondary rural-urban commuting area code &  &  &  & 0.798 \\ 
      -  1 & 673 (49.3\%) & 556 (45.9\%) & 1229 (47.7\%) &  \\ 
      -  2 & 259 (19.0\%) & 246 (20.3\%) & 505 (19.6\%) &  \\ 
      -  2.1 & 3 (0.2\%) & 1 (0.1\%) & 4 (0.2\%) &  \\ 
      -  3 & 22 (1.6\%) & 17 (1.4\%) & 39 (1.5\%) &  \\ 
      -  4 & 144 (10.5\%) & 143 (11.8\%) & 287 (11.1\%) &  \\ 
      -  4.1 & 0 (0.0\%) & 1 (0.1\%) & 1 (0.0\%) &  \\ 
      -  5 & 55 (4.0\%) & 49 (4.0\%) & 104 (4.0\%) &  \\ 
      -  5.1 & 1 (0.1\%) & 0 (0.0\%) & 1 (0.0\%) &  \\ 
      -  6 & 21 (1.5\%) & 22 (1.8\%) & 43 (1.7\%) &  \\ 
      -  7 & 54 (4.0\%) & 62 (5.1\%) & 116 (4.5\%) &  \\ 
      -  7.1 & 34 (2.5\%) & 32 (2.6\%) & 66 (2.6\%) &  \\ 
      -  8 & 37 (2.7\%) & 30 (2.5\%) & 67 (2.6\%) &  \\ 
      -  9 & 41 (3.0\%) & 37 (3.1\%) & 78 (3.0\%) &  \\ 
      -  10 & 21 (1.5\%) & 16 (1.3\%) & 37 (1.4\%) &  \\
      Rural-urban continuum code &  &  &  & 0.235 \\ 
      -  1 & 528 (38.7\%) & 459 (37.9\%) & 987 (38.3\%) &  \\ 
      -  2 & 296 (21.7\%) & 229 (18.9\%) & 525 (20.4\%) &  \\ 
      -  3 & 218 (16.0\%) & 244 (20.1\%) & 462 (17.9\%) &  \\ 
      -  4 & 196 (14.4\%) & 173 (14.3\%) & 369 (14.3\%) &  \\ 
      -  5 & 0 (0.0\%) & 1 (0.1\%) & 1 (0.0\%) &  \\ 
      -  6 & 110 (8.1\%) & 94 (7.8\%) & 204 (7.9\%) &  \\ 
      -  7 & 2 (0.1\%) & 1 (0.1\%) & 3 (0.1\%) &  \\ 
      -  8 & 12 (0.9\%) & 9 (0.7\%) & 21 (0.8\%) &  \\ 
      -  9 & 3 (0.2\%) & 2 (0.2\%) & 5 (0.2\%) &  \\ 
      Median Age &  &  &  & 0.242 \\ 
      -  Mean (SD) & 40.092 (6.488) & 39.804 (5.979) & 39.957 (6.254) &  \\ 
      -  Range & 21.300 - 64.400 & 22.000 - 64.400 & 21.300 - 64.400 &  \\ 
      Total population (2015) &  &  &  & 0.496 \\ 
      -  Mean (SD) & 5947.348 (2428.290) & 6011.554 (2340.613) & 5977.545 (2387.209) &  \\ 
      -  Range & 1327.000 - 16788.000 & 1673.000 - 16788.000 & 1327.000 - 16788.000 &  \\ 
      Median income (USD) &  &  &  & $<$ 0.001 \\ 
      -  Mean (SD) & 56310.915 (22716.651) & 52997.941 (21256.845) & 54752.776 (22099.857) &  \\ 
      -  Range & 13456.000 - 160000.000 & 13456.000 - 151875.000 & 13456.000 - 160000.000 &  \\ 
      Average household size &  &  &  & 0.073 \\ 
      -  Mean (SD) & 2.557 (0.307) & 2.578 (0.294) & 2.567 (0.301) &  \\ 
      -  Range & 1.410 - 3.490 & 1.720 - 3.400 & 1.410 - 3.490 &  \\ 
      Median monthly rent (USD) &  &  &  & $<$ 0.001 \\ 
      -  Mean (SD) & 850.325 (227.908) & 819.452 (214.420) & 835.805 (222.159) &  \\ 
      -  Range & 458.000 - 2196.000 & 435.000 - 2196.000 & 435.000 - 2196.000 &  \\ 
      Median home value (USD) &  &  &  & $<$ 0.001 \\ 
      -  Mean (SD) & 188344.322 (103522.187) & 172789.274 (93681.192) & 181028.560 (99300.824) &  \\ 
      -  Range & 50600.000 - 701500.000 & 50300.000 - 654100.000 & 50300.000 - 701500.000 &  \\ 
      \% Over 85 &  &  &  & 0.252 \\ 
      -  Mean (SD) & 1.334 (1.911) & 1.253 (1.604) & 1.296 (1.773) &  \\ 
      -  Range & 0.000 - 11.600 & 0.000 - 11.600 & 0.000 - 11.600 &  \\
      \% Over 65 &  &  &  & 0.413 \\ 
      -  Mean (SD) & 38.508 (6.649) & 38.300 (6.169) & 38.410 (6.427) &  \\ 
      -  Range & 21.300 - 62.000 & 18.900 - 62.000 & 18.900 - 62.000 &  \\
     \& With disability &  &  &  & 0.006 \\ 
      -  Mean (SD) & 13.427 (5.590) & 14.019 (5.348) & 13.705 (5.485) &  \\ 
      -  Range & 1.800 - 34.100 & 1.800 - 34.100 & 1.800 - 34.100 &  \\ 
      \% Receiving SNAP/food stamp benefits &  &  &  & $<$ 0.001 \\ 
      -  Mean (SD) & 13.381 (9.615) & 14.669 (9.883) & 13.987 (9.761) &  \\ 
      -  Range & 0.000 - 63.500 & 0.000 - 63.500 & 0.000 - 63.500 &  \\ 
      \% Public health insurance with medicare &  &  &  & $<$ 0.001 \\ 
      -  Mean (SD) & 16.949 (8.536) & 18.262 (8.923) & 17.566 (8.743) &  \\ 
      -  Range & 1.000 - 49.400 & 0.800 - 60.200 & 0.800 - 60.200 &  \\ 
      \% Public health insurance with medicaid &  &  &  & 0.031 \\ 
      -  Mean (SD) & 52.188 (16.387) & 53.542 (15.371) & 52.824 (15.929) &  \\ 
      -  Range & 0.000 - 100.000 & 0.000 - 100.000 & 0.000 - 100.000 &  \\ 
      \% Public health insurance with VA health care &  &  &  & 0.988 \\ 
      -  Mean (SD) & 85.259 (20.999) & 85.247 (20.842) & 85.254 (20.921) &  \\ 
      -  Range & 0.000 - 100.000 & 0.000 - 100.000 & 0.000 - 100.000 &  \\ 
     \% Bachelor &  &  &  & $<$ 0.001 \\ 
      -  Mean (SD) & 33.532 (21.945) & 29.999 (20.673) & 31.870 (21.425) &  \\ 
      -  Range & 2.700 - 84.300 & 2.700 - 84.500 & 2.700 - 84.500 &  \\ 
      \% Foreign born &  &  &  & 0.745 \\ 
      -  Mean (SD) & 8.321 (6.096) & 8.400 (6.163) & 8.358 (6.126) &  \\ 
      -  Range & 0.000 - 32.700 & 0.000 - 42.000 & 0.000 - 42.000 &  \\ 
      \% Unemployed &  &  &  & 0.002 \\ 
      -  Mean (SD) & 8.565 (4.331) & 9.106 (4.665) & 8.820 (4.498) &  \\ 
      -  Range & 0.200 - 29.500 & 0.500 - 30.200 & 0.200 - 30.200 &  \\ 
      \% Commute with carpool &  &  &  & 0.090 \\ 
      -  Mean (SD) & 10.114 (5.030) & 10.458 (5.259) & 10.275 (5.141) &  \\ 
      -  Range & 0.500 - 32.800 & 0.500 - 32.800 & 0.500 - 32.800 &  \\ 
      Average commute time (min) &  &  &  & 0.611 \\ 
      -  Mean (SD) & 24.606 (4.857) & 24.703 (4.891) & 24.652 (4.872) &  \\ 
      -  Range & 13.700 - 37.600 & 13.300 - 40.100 & 13.300 - 40.100 &  \\
      \% Working in agriculture &  &  &  & 0.057 \\ 
      -  Mean (SD) & 2.302 (4.124) & 2.625 (4.489) & 2.454 (4.302) &  \\ 
      -  Range & 0.000 - 26.900 & 0.000 - 26.900 & 0.000 - 26.900 &  \\ 
      \% Working in construction &  &  &  & 0.039 \\ 
      -  Mean (SD) & 6.833 (3.991) & 7.162 (4.116) & 6.988 (4.053) &  \\ 
      -  Range & 0.000 - 25.600 & 0.000 - 22.600 & 0.000 - 25.600 &  \\ 
      \% Working in manufacturing &  &  &  & 0.014 \\ 
      -  Mean (SD) & 11.896 (5.644) & 12.453 (5.845) & 12.158 (5.745) &  \\ 
      -  Range & 1.100 - 39.100 & 1.600 - 42.600 & 1.100 - 42.600 &  \\ 
      \% Working in wholesale trade &  &  &  & 0.457 \\ 
      -  Mean (SD) & 3.039 (1.892) & 3.095 (1.902) & 3.066 (1.896) &  \\ 
      -  Range & 0.000 - 11.300 & 0.000 - 11.300 & 0.000 - 11.300 &  \\ 
      \% Working in retail trade &  &  &  & 0.086 \\ 
      -  Mean (SD) & 11.222 (3.814) & 11.484 (3.910) & 11.345 (3.861) &  \\ 
      -  Range & 2.300 - 31.500 & 2.300 - 31.500 & 2.300 - 31.500 &  \\ 
      \% Working in transportation, warehousing and utilities &  &  &  & 0.058 \\ 
      -  Mean (SD) & 3.509 (2.280) & 3.677 (2.199) & 3.588 (2.243) &  \\ 
      -  Range & 0.000 - 12.700 & 0.000 - 11.800 & 0.000 - 12.700 &  \\ 
      \% Working in information &  &  &  & 0.214 \\ 
      -  Mean (SD) & 1.778 (1.476) & 1.706 (1.456) & 1.744 (1.467) &  \\ 
      -  Range & 0.000 - 11.300 & 0.000 - 8.900 & 0.000 - 11.300 &  \\ 
      \% Working in finance, insurance and real estate &  &  &  & 0.016 \\ 
      -  Mean (SD) & 5.497 (2.946) & 5.219 (2.874) & 5.366 (2.915) &  \\ 
      -  Range & 0.000 - 20.100 & 0.000 - 13.600 & 0.000 - 20.100 &  \\ 
      \% Working in professional, sicentific, management, administrative and waste management &  &  &  & 0.001 \\ 
      -  Mean (SD) & 11.144 (6.523) & 10.341 (6.182) & 10.767 (6.376) &  \\ 
      -  Range & 0.800 - 30.500 & 1.200 - 30.500 & 0.800 - 30.500 &  \\ 
      \% Working in education, health care and social assistance &  &  &  & 0.020 \\ 
      -  Mean (SD) & 24.859 (7.473) & 24.182 (7.193) & 24.540 (7.349) &  \\ 
      -  Range & 4.500 - 51.600 & 9.900 - 56.500 & 4.500 - 56.500 &  \\ 
      \% Working in arts and entertainment &  &  &  & 0.734 \\ 
      -  Mean (SD) & 8.083 (3.676) & 8.033 (3.721) & 8.060 (3.696) &  \\ 
      -  Range & 0.300 - 27.700 & 0.300 - 28.700 & 0.300 - 28.700 &  \\ 
      \% Working in other services &  &  &  & 0.011 \\ 
      -  Mean (SD) & 4.595 (2.108) & 4.808 (2.134) & 4.695 (2.123) &  \\ 
      -  Range & 0.000 - 11.400 & 0.000 - 14.800 & 0.000 - 14.800 &  \\ 
      \% Working in public administration &  &  &  & 0.831 \\ 
      -  Mean (SD) & 5.231 (2.797) & 5.208 (2.766) & 5.220 (2.782) &  \\ 
      -  Range & 0.000 - 18.300 & 0.000 - 20.500 & 0.000 - 20.500 &  \\ 
      \% Private health insurance &  &  &  & $<$ 0.001 \\ 
      -  Mean (SD) & 67.509 (15.111) & 65.148 (15.644) & 66.398 (15.406) &  \\ 
      -  Range & 20.900 - 97.600 & 17.200 - 95.600 & 17.200 - 97.600 &  \\ 
      \% Public health insurance &  &  &  & 0.018 \\ 
      -  Mean (SD) & 32.444 (11.161) & 33.477 (11.037) & 32.930 (11.113) &  \\ 
      -  Range & 5.300 - 67.000 & 6.300 - 72.600 & 5.300 - 72.600 &  \\ 
      \% No health insurance &  &  &  & $<$ 0.001 \\ 
      -  Mean (SD) & 13.038 (6.434) & 13.893 (6.614) & 13.440 (6.532) &  \\ 
      -  Range & 0.500 - 37.700 & 1.100 - 36.000 & 0.500 - 37.700 &  \\ 
      \% White &  &  &  & 0.169 \\ 
      -  Mean (SD) & 69.123 (18.016) & 68.157 (17.526) & 68.669 (17.790) &  \\ 
      -  Range & 3.706 - 98.928 & 3.706 - 99.021 & 3.706 - 99.021 &  \\ 
      \% Black &  &  &  & 0.160 \\ 
      -  Mean (SD) & 21.511 (16.761) & 22.427 (16.239) & 21.942 (16.521) & \\
      -  Range & 0.000 - 95.134 & 0.000 - 95.134 & 0.000 - 95.134 &  \\ 
      \% American Indian &  &  &  & 0.187 \\ 
      -  Mean (SD) & 0.929 (4.120) & 0.748 (2.531) & 0.844 (3.465) &  \\ 
      -  Range & 0.000 - 78.041 & 0.000 - 51.952 & 0.000 - 78.041 &  \\ 
      \% Asian &  &  &  & 0.050 \\ 
      -  Mean (SD) & 2.600 (4.587) & 2.253 (4.375) & 2.436 (4.491) &  \\ 
      -  Range & 0.000 - 35.790 & 0.000 - 48.531 & 0.000 - 48.531 &  \\ 
      \% Pacific Islander &  &  &  & 0.005 \\ 
      -  Mean (SD) & 0.054 (0.274) & 0.030 (0.131) & 0.043 (0.219) &  \\ 
      -  Range & 0.000 - 5.515 & 0.000 - 1.636 & 0.000 - 5.515 &  \\ 
      \% Other race &  &  &  & 0.003 \\ 
      -  Mean (SD) & 3.475 (4.687) & 4.062 (5.464) & 3.751 (5.074) &  \\ 
      -  Range & 0.000 - 37.124 & 0.000 - 37.124 & 0.000 - 37.124 &  \\ 
      \% Mixed race &  &  &  & 0.836 \\ 
      -  Mean (SD) & 2.309 (1.712) & 2.323 (1.672) & 2.316 (1.693) &  \\ 
      -  Range & 0.000 - 10.328 & 0.000 - 10.017 & 0.000 - 10.328 &  \\ 
      \% Hispanic/Latino &  &  &  & 0.004 \\ 
      -  Mean (SD) & 9.400 (8.238) & 10.401 (9.326) & 9.871 (8.779) &  \\ 
      -  Range & 0.000 - 48.524 & 0.000 - 48.524 & 0.000 - 48.524 &  \\ 
      \% Living in owned unit &  &  &  & 0.043 \\ 
      -  Mean (SD) & 67.463 (15.424) & 66.238 (15.267) & 66.887 (15.359) &  \\ 
      -  Range & 14.215 - 97.896 & 4.616 - 96.298 & 4.616 - 97.896 &  \\ 
      \% Living in rented unit &  &  &  & 0.036 \\ 
      -  Mean (SD) & 30.944 (14.604) & 32.146 (14.406) & 31.509 (14.521) &  \\ 
      -  Range & 2.104 - 82.024 & 3.702 - 88.201 & 2.104 - 88.201 &  \\ 
      \% Veteran &  &  &  & 0.495 \\ 
      -  Mean (SD) & 7.140 (2.694) & 7.068 (2.585) & 7.106 (2.643) &  \\ 
      -  Range & 0.558 - 22.024 & 1.295 - 22.024 & 0.558 - 22.024 &  \\ 

\end{longtable}

\end{landscape}

\end{document}